\documentclass{iopart}
\usepackage{graphicx,epsfig}
\usepackage{bm}
\usepackage{iopams}
\newcommand{\ba}{\begin{eqnarray}}
\newcommand{\ea}{\end{eqnarray}}
\newcommand{\bse}{\numparts}
\newcommand{\ese}{\endnumparts}

\newcommand{\DD}{{\cal {D}}}

\newcommand{\bbq}{\begin{quote}}
\newcommand{\eeq}{\end{quote}}
\newcommand{\RR}{{}^3{\cal{R}}}

\newcommand{\T}{{}^3{\cal{T}}}

\newcommand{\EE}{{\cal{E}}}

\newcommand{\PP}{{\cal{P}}}

\newcommand{\Oml}{\hat\Omega^{(\lambda)}}
\newcommand{\Omli}{\hat\Omega_i^{(\lambda)}}
\newcommand{\Omm}{\hat\Omega^{(\mu)}}
\newcommand{\Ommi}{\hat\Omega_i^{(\mu)}}
\newcommand{\Da}{\delta^{(A)}}

\newcommand{\Dim}{\delta_i^{(\mu)}}
\newcommand{\Dik}{\delta_i^{(\kappa)}}
\newcommand{\Diz}{\delta_i^{(z)}}

\newcommand{\Dm}{\delta^{(\mu)}}

\newcommand{\Dk}{\delta^{(\kappa)}}

\newcommand{\Dz}{\delta^{(z)}}

\newcommand{\dd}{{\rm{d}}}
\newcommand{\Qmin}{Q_{\textrm{\tiny{min}}}}
\newcommand{\Lmin}{L_{\textrm{\tiny{min}}}}
\newcommand{\rtv}{r_{\textrm{\tiny{tv}}}}

\begin{document}


\title[A dynamical systems study of the inhomogeneous $\Lambda$ CDM model]{A dynamical systems study of the inhomogeneous $\Lambda$ CDM model.}
\author{ Roberto A. Sussman${}^\dagger$ and Germ\'an Izquierdo${}^\ddagger$}
\address{${}^\dagger$ Instituto de Ciencias Nucleares, Universidad Nacional Aut\'onoma de M\'exico (ICN-UNAM),A. P. 70--543, 04510 M\'exico D. F., M\'exico.\\
 ${}^\ddagger$ Institut de Math\'ematiques de Bourgogne, Universit\'e de Bourgogne, UFR Sciences et Techniques, Facult\'e des Sciences Mirande, 9 Av. Alain Savary, 21078 Dijon Cedex, France}
\ead{sussman@nucleares.unam.mx, german.izquierdo@gmail.com}
\date{\today}
\begin{abstract} We consider spherically symmetric inhomogeneous dust models with a positive cosmological constant, $\Lambda$, given by the Lema\^\i tre--Tolman--Bondi metric. These configurations provide a simple but useful generalization of the $\Lambda$--CDM model describing cold dark matter (CDM) and a $\Lambda$ term, which seems to fit current cosmological observations. The dynamics of these models can be fully described by scalar evolution equations that can be given in the form of a proper dynamical system associated with a 4--dimensional phase space whose critical points and invariant subspaces are examined and classified. The phase space evolution of various configurations is studied in detail by means of two 2--dimensional subspaces: a projection into the invariant homogeneous subspace associated with $\Lambda$--CDM solutions with FLRW metric, and a projection into a subspace generated by suitably defined fluctuations that convey the effects of inhomogeneity. We look at cases with perpetual expansion, bouncing and loitering behavior, as well as configurations with ``mixed'' kinematic patters, such as a collapsing region in an expanding background. In all cases, phase space trajectories emerge from and converge to stable past and future attractors  in a qualitatively analogous way as in the case of the FLRW limit. However, we can identify in both projections of the phase space various qualitative features absent in the FLRW limit that can be useful in the construction of toy models of astrophysical and cosmological inhomogeneities.
\end{abstract}

\section{Introduction.}

From a phenomenological and empiric point of view, cosmological observations are usually tested as a first attempt by fitting them to the so--called $\Lambda$--CDM model, which is a FLRW spacetime with flat spacelike sections, whose source is dust (cold dark matter CDM) and a $\Lambda$ field (dark energy) (see \cite{reviewDE1,reviewDE2} for comprehensive reviews). Although the universe may be nearly homogeneous at scales larger than the so--called homogeneity scale (over 150--300 Mpc), thus justifying the use of linear perturbations in its dynamical study, it is clearly inhomogeneous at smaller scales in which structure formation mostly involving CDM has taken  place. Hence, the study of inhomogeneous sources made of dust and a nonzero ``$\Lambda$ field'' is a very relevant topic. In particular, spacetimes of this type of source with spherical symmetry provide simple but non--trivial inhomogeneous generalizations of the $\Lambda$--CDM model.

Inhomogeneous spherically symmetric dust solutions of Einstein's equations with $\Lambda=0$,  described by the Lema\^\i tre--Tolman--Bondi (LTB) metric have been widely used to construct simple models of cosmological inhomogeneities (see \cite{kras1,kras2} for a comprehensive discussion and review). However, the literature on LTB models with $\Lambda> 0$ (which we will denote by $\Lambda$--LTB models) is much less abundant \cite{LTBL1,LTBL2}, dealing mostly on singularities and censorship \cite{LTBLS}. See \cite{WA} for a recent study of the dynamics of LTB models (considering also $\Lambda\ne 0$), while Szekeres models with $\Lambda\ne 0$ have been examined in \cite{barrow}.

In the present paper we attempt to study the $\Lambda$--LTB models under the qualitative and numerical phase space techniques known as ``dynamical systems''. These techniques have been applied successfully to a wide variety of spacetimes (FLRW, Kantowski--Sachs, Bianchi models, self--similar and static spacetimes, etc), for which evolution equations for the ``expansion normalized'' phase variables can be reduced to a system of autonomous ODE's that define a self--consistent phase space (see \cite{DS1,DS2,DS3,DS4} for a comprehensive elaboration and review). A dynamical system analysis has been been successfully applied to LTB dust solutions with $\Lambda=0$ in reference \cite{suss08}. We generalize here this study to the case $\Lambda> 0$. The plan of this article is summarized below.

The class of $\Lambda$--LTB spacetimes and their corresponding covariant objects are introduced in section 2, while in section 3 we show that their dynamics can be completely determined by ``quasi--local'' (QL) variables \cite{sussPRD,sussLTB1,sussLTB2} that lead to a set of ``fluid flow'' evolution equations \cite{1plus3} and an initial value parametrization that is useful for carrying on numeric work on the models. These evolution equations  are transformed in section 4 into a 4--dimensional dynamical system constructed by ``expansion normalized'' variables  \cite{DS1,DS2,DS3,DS4}. In section 5 we list the critical points (past and future attractors and saddle points) of this dynamical system and classify all the  invariant subspaces of the phase space: the homogeneous FLRW subset, the spatially flat subspace, as well as the Kottler vacuum solution \cite{kottler} (Schwarzschild--de Sitter) and ``pure'' dust with $\Lambda=0$ subspaces. We examine in section 6 the different patterns of kinematic evolution of $\Lambda$--LTB models. The admissible topological classes of the space slices orthogonal to the 4--velocity are summarized in section 7. Since the phase space is 4--dimensional, we decompose it in section 8 in two 2--dimensional projections: the ``homogeneous'' and ``inhomogeneous'' subsystems. These projections convey the full dynamical information, and thus are used in section 9 to study in detail the phase space trajectories of representative configurations associated with each one of the kinematic patterns and topologies of the space slices listed in sections 6 and 7. For this purpose, we solve numerically the dynamical system obtained in section 4 for each configuration, together with the QL evolution equations derived in section 3 (which is necessary when the phase space variables diverge in collapsing or bouncing configurations). Section 10 summarizes the results and provides a final discussion.

\section{$\Lambda$--LTB spacetimes}\label{genLTB}

Spherically symmetric dust solutions with nonzero cosmological constant, $\Lambda$, or (as we shall call them) ``$\Lambda$--Lema\^\i tre--Tolman--Bondi" ($\Lambda$--LTB) models are characterized by the LTB metric \cite{kras1,kras2}
\begin{equation} ds^2 = -c^2dt^2 +\frac{R'^2\,dr^2}{1+E}+R^2[d\theta^2+\sin^2\theta\,d\phi^2],\label{LTB}\end{equation}
where $R=R(x^0,r)$, \, ($x^0=ct$),\, \, $R'=\partial R/\partial r$,\, $E=E(r)$, and the energy--momentum tensor
\begin{equation}T^{ab} = \rho\,u^a u^b -\Lambda\,g^{ab},\label{Tab}\end{equation}
where $u^a=\delta^a_0$ and $\rho(x^0,r)$ is the rest--mass density. The field equations $G^{ab}=\kappa T^{ab}$ (with $\kappa_0 =8\pi G/c^4$) for (\ref{Tab})--(\ref{LTB}) yield:
\ba \kappa_0 \rho c^2 R^2R' = 2M',\label{rho_dust}
\\ \dot R^2 = \frac{2M}{R}+\frac{\kappa_0}{3}\Lambda R^2+E,\label{fried_dust}
\ea
where where $\dot R=\partial R/\partial x^0$, and the function $M=M(r)$, which appears as an ``integration constant', is the ``effective'' rest mass--energy in length units.

The comoving geodesic 4--velocity field defines a foliation of spacelike hypersurfaces, $\T(t)$, orthogonal to $u^a$, marked by $t$ constant, and having an induced metric $h_{ab}=u_au_b+g_{ab}=\delta_a^i\delta_b^jg_{ij}$, where $i,j = r,\theta,\phi$. Besides $u^a,\,\rho$ and $\Lambda$, the remaining covariant objects associated with $\Lambda$--LTB spacetimes are two scalars: the expansion, $\Theta=\nabla_a u^a$, and $\RR$, the Ricci scalar of hypersurfaces $\T(t)$:
\begin{equation} \Theta = \frac{2\dot R}{R}+\frac{\dot R'}{R'},\qquad 
\RR  = -\frac{2(ER)'}{R^2R'},\label{Theta1RR1} \end{equation}
plus two spacelike traceless tensors, the shear tensor, $\sigma_{ab}=\nabla_{(a}u_{b)}-(\Theta/3)\,h_{ab}$, and the electric Weyl tensor, $E_{ab}=C_{acbd}u^cu^d$. Both of these tensors can be completely and covariantly determined in terms of a single scalar function as $\sigma^{ab} =\Sigma\,\Xi^{ab}$ and $E^{ab} = \EE\,\Xi^{ab}$, where $\Xi^{ab}\equiv h^{ab}-3\chi^{a}\chi^b$\, with $\chi^a=\sqrt{h^{rr}}\,\delta^a_r$ being the unit vector orthogonal to $u^a$ and to the orbits of SO(3). The functions $\Sigma$ and $\EE$ are 
\begin{equation} \Sigma = \frac{1}{3}\left[\frac{\dot R}{R}-\frac{\dot R'}{R'}\right],\qquad
\EE = -\frac{\kappa_0}{6}\,\rho\,c^2+\frac{M}{R^3}.\label{Sigma1}\end{equation}
The set $\{\rho,\,\Lambda,\,\Theta,\,\Sigma,\,\EE,\,\RR\}$ then provides a complete representation of local covariant scalars for the $\Lambda$--LTB models.

\section{Quasi--local (QL) scalar variables.} \label{QLdefs}

Following~\cite{sussPRD,sussLTB1}, we define for the scalars $A=\{\rho,\,\Theta,\,\RR,\,\Lambda\}$ their dual quasi---local QL scalars $A_q=\{\rho_q,\,\Theta_q,\,\RR_q,\,\Lambda_q\}$ by means of the integral
\footnote{We are using a different notation from that of \cite{sussPRD,sussLTB1}, where QL functions $A_q$ were denoted by $A_*$. These same functions where introduced in \cite{suss08}, but were not called QL functions. In that reference the notation used was:\, $\ell,\,H,\,S$ respectively for $L,\,\Theta_q/3,\,-\Dz$ and $\Delta^{(m)},\,\Delta^{(k)}$ for $\Dm,\,\Dk$, while $\langle \rho\rangle,\, \langle \RR\rangle$ and $\langle\Omega\rangle$ were used for $\rho_q,\,\RR_q$ and $\Omm$. The use of the notation $\langle..\rangle$ can be misleading because (\ref{QLfunc}) is not an average unless it is defined as a functional (see \cite{sussBR}). See also  \cite{sussPRD,sussLTB1,hayward} for a discussion of the covariant nature of the integral (\ref{QLfunc}) and its relation to the Misner--Sharp QL mass--energy function.}
\begin{equation}A_q = \frac{\int_{x=0}^{x=r}{A\,R^2R' dx}}{\int_{x=0}^{x=r}{R^2R' dx}}.\label{QLfunc}\end{equation}
where the integral is evaluated in spherical comoving domains $\DD=\mathbb{S}^2\times\vartheta[r]\subset \T(t)$, where $\mathbb{S}^2$ is the unit 2--sphere parametrized by $(\theta,\phi)$ and $\vartheta[r] =\{x\,|\, 0\leq x\leq r\}$, where $x=0$ marks a symmetry center. It is evident that $\Lambda_q=\Lambda$.
\footnote{The radial range can be modified to include models not admitting a symmetry center or when the $\T[t]$ are restricted by curvature singularities, see \cite{sussPRD,sussLTB1,sussLTB2}}
 From (\ref{Theta1RR1}) and (\ref{QLfunc}), the QL scalars are
\begin{equation}\fl \frac{\kappa_0}{3}\,\rho_q = \frac{2M}{R^3},\qquad \frac{\Theta_q}{3}=\frac{\dot R}{R},\qquad  \frac{\RR_q}{6} = -\frac{E}{R^2},\qquad \Lambda_q =\Lambda.\label{qvars}\end{equation}
The local and QL scalars can be related by means of ``relative deviations''
\begin{equation} \Da \equiv \frac{A-A_q}{A_q}=\frac{A'_q/A_q}{3R'/R}\label{Da_def}\end{equation}
where we used the property: $A'_q=(3R'/R)(A-A_q)$ which follows directly from (\ref{QLfunc}).

It is convenient for numerical work to use dimensionless variables, hence we introduce
\begin{equation}\fl 2\mu_q \equiv \frac{\kappa_0\,\rho_q c^2}{3h_s^2},\quad z_q\equiv \frac{\Theta_q}{3h_s},\quad \kappa_q\equiv \frac{\RR_q}{6\,h_s^2},\quad \lambda\equiv \frac{\kappa_0\,\Lambda}{3h_s^2},\quad \tau = h_s\,c(t-t_i),\label{dimless}\end{equation}
where $h_s^{-1}$ is an arbitrary length scale and $t=t_i$ marks a fiducial initial hypersurface $\T_i$, so that it coincides with $\tau=0$. Considering the definitions (\ref{dimless}) and following \cite{sussPRD,sussLTB1}, the dynamics of LTB-dS models is fully determined by the following evolution equations
\footnote{Equation (\ref{sys1d}) was given incorrectly in \cite{sussPRD,sussLTB1}.}
\bse\ba \frac{\partial\, \mu_q}{\partial \tau} &=& -3\,\mu_q\,z_q,\label{sys1a}\\
\frac{\partial \,z_q}{\partial \tau} &=& -z_q^2-\mu_q+\lambda,\label{sys1b}\\
\frac{\partial \,\Dm}{\partial \tau} &=& -3\,z_q\,\Dz\,\left[1+\Dm\right],\label{sys1c}\\
\frac{\partial \,\Dz}{\partial \tau} &=& -z_q\,\delta^{(z)}\,\left[1+3\Dz\right]+
\frac{\mu_q\,\left[\Dz-\Dm\right]-\lambda\,\Dz}{z_q}.\label{sys1d}\ea\ese
together with the constraint
\begin{equation} z_q^2 = 2\mu_q -\kappa_q+\lambda,\label{Hconst}\end{equation}
Once the system (\ref{sys1a})--(\ref{sys1d}) is solved, all scalars $\{\rho,\,\Lambda,\,\Theta,\,\Sigma,\,\EE,\,\RR\}$ follow as:
\bse\ba\fl \mu=\mu_q\,\left[1+\Dm\right],\qquad z=z_q\,\left[1+\Dz\right],\qquad \kappa=\kappa_q\,\left[1+\Dk\right],\label{mzk}\\
\fl \Sigma = -h_s z_q\Dz,\qquad \EE = -h_s \mu_q\Dm,\label{SE}\ea\ese
As proven in \cite{sussPRD}, evolution equations like (\ref{sys1a})--(\ref{sys1d}) completely determine the dynamics of all spacetimes described by an LTB metric in the comoving frame (including $\Lambda$--LTB models). In this description the variables $A_q$ (which satisfy FLRW dynamics) formally define a FLRW background, while the $\Da$ can be rigorously characterized as covariant, gauge--invariant and non--linear perturbations on this background.

It is useful to introduce an initial value formulation for the LTB--dS models.  By considering an arbitrary slice $\T_i=\T[t_i]$, we rephrase the metric functions $R$ and $R'$ in (\ref{LTB}) as dimensionless scale factors
\begin{equation} L = \frac{R}{R_i},\qquad
 \Gamma = \frac{R'/R}{R'_i/R_i}=1+\frac{L'/L}{R'_i/R_i}.\label{LGdef}\end{equation}
where the subindex ${}_i$ will denote henceforth evaluation at $t=t_i$. Considering as initial conditions for (\ref{sys1a})--(\ref{sys1d}) the initial value functions $\{\mu_{qi},\, \kappa_{qi},\,\lambda\}$, the standard free parameters in (\ref{LTB})--(\ref{rho_dust}) follow from (\ref{qvars}) as $M  = \mu_{qi}\,h_s^{2}\,R_i^3$ and $E  = -\kappa_{qi}\,h_s^{2}\,R_i^2$, leading from (\ref{fried_dust}), (\ref{qvars}) and (\ref{mzk}) to the Friedman--like equation:
\begin{equation} z_q^2L^2=L_{,\tau}^2 = \frac{Q(L)}{L},\qquad Q(L)\equiv 2\mu_{qi}-\kappa_{qi}\,L+\lambda\,L^3\label{sqHq}\end{equation}
and the following scaling laws
\ba \mu_q=\frac{\mu_{qi}}{L^3},\qquad \kappa_q=\frac{\kappa_{qi}}{L^2},\qquad z_q=\frac{L_{,\tau}}{L},\label{slaw_mkH}\\
 1+\Dm =\frac{1+\Dim}{\Gamma},\qquad \frac{2}{3}+\Dk = \frac{2/3+\Dik}{\Gamma},\label{slaw_Dmk}\\
2\Dz = \Omm\,\Dm + \left[1-\Omm-\Oml\right]\,\Dk,\label{slaw_Dh}\ea
where the ``Omega'' factors are
\ba \Omm \equiv \frac{\kappa_0\,\rho_q\,c^2}{3\,h_s^2\,z_q^2}=\frac{2\mu_q}{z_q^2}=\frac{2\mu_{qi}}{Q(L)},\label{Omm}
\\
 \Oml \equiv \frac{\kappa_0\,\Lambda}{3\,h_s^2\,z_q^2}=\frac{\lambda}{z_q^2}=\frac{2\lambda\,L^3}{Q(L)},\label{Oml}\ea
The metric (\ref{LTB}) in this initial value formulation takes the following FLRW--like form:
\footnote{The metric (\ref{LTB2}) simplifies with the coordinate choice (\ref{eqRi}) in section 7, leaving $h_s^{-1}$ as the only characteristic length.}
\begin{equation} \dd s^2=-c^2\dd t^2+L^2\left[\frac{\Gamma^2\,R'_i{}^2\,\dd r^2}{1-\kappa_{qi}\,h_s^2\, R_i^2}+R_i^2\left(\dd\theta^2+\sin^2\theta\dd \phi^2\right)\right],\label{LTB2}\end{equation}
so that $L(t,r)=0$ and $\Gamma(t,r)=0$ respectively mark the central and shell crossing  singularities. The scale factor $L$ follows from the implicit solutions of (\ref{sqHq}), which as opposed to the case $\lambda=0$, cannot be given (in general) in terms of elementary functions, but by elliptic integrals that can be formally written as~\cite{LTBL1,LTBL2}
\begin{equation} \fl \tau = Z-Z_i,\quad Z=Z(L,\mu_{qi},\kappa_{qi},\lambda)=\int{\frac{L^{1/2}\,\dd L}{\left[2\mu_{qi}+\lambda L^3-\kappa_{qi} L\right]^{1/2}}},\label{Zdef}\end{equation}
Hence, a formal expression for $\Gamma$ can be obtained by implicit radial derivation of (\ref{Zdef}):
\begin{equation} \fl\Gamma = 1- \frac{3[\mu_{qi}\,(Z-Z_i)_{,\mu_{qi}}\,\Dim+\kappa_{qi}\,(Z-Z_i)_{,\kappa_{qi}}\,\Dik]}{L\,Z_{,L}},\label{Gamma}\end{equation}
where the subindices ${}_{,\mu_{qi}}$ and ${}_{,\kappa_{qi}}$ denote partial derivatives with respect to $\mu_{qi}$ and $\kappa_{qi}$, while $Z_i$ is the elliptic integral $Z$ evaluated at $\tau=0$ ($L=1$). As a consequence, the conditions to avoid shell crossings, $\Gamma>0$, cannot be (in general) given analytically as in the case $\lambda=0$ (see \cite{ltbstuff,suss02}).

\section{A dynamical system.}

The evolution equations (\ref{sys1a})--(\ref{sys1d}) lead in a natural manner to a dynamical system based on ``expansion normalized'' variables~\cite{DS1,DS2,DS3,DS4}. Following the methodology of \cite{suss08}, we define the dynamical system  variables by normalizing with the QL expansion scalar,  $z_q$, and by considering the ``Omega'' parameters given by (\ref{Omm})--(\ref{Oml}). We define also a suitable evolution parameter $\xi$ by means of the coordinate transformation
\begin{equation} \tau =\tau(\xi,\bar r),\qquad r=\bar r,\end{equation}
so that for every scalar $A(\tau,r)=A(\tau(\xi,r),r)=A(\xi,r)$. Choosing $\xi(\tau,r)=\ln (L)$, so that $[\partial\xi/\partial\tau]_r=z_q=h_s^{-1}\dot R/R$, leads for comoving curves ($r=$ constant) to the differential operator
\begin{equation}\frac{\partial}{\partial\xi}=\frac{1}{z_q}\frac{\partial}{\partial\tau}
=\frac{3}{\Theta_q}\frac{\partial}{c\partial t}.\label{xidef}\end{equation}
Eliminating $\mu_q$ and $\lambda$ in terms of $\Omm$ and $\Oml$ from (\ref{Omm})--(\ref{Oml}) and using (\ref{xidef}), the system (\ref{sys1a})--(\ref{sys1d}) transforms into the following dynamical system:
\bse\ba \frac{\partial\,\Omm }{\partial\xi} &=& \Omm\,\left[\Omm-2\Oml-1\right],\label{DSa}\\
\frac{\partial\,\Oml }{\partial\xi} &=& \Oml\,\left[2\left(1-\Oml\right)+\Omm\right],\label{DSb}\\
\frac{\partial\,\Dm }{\partial\xi} &=& -3\,\left[1+\Dm\right]\,\Dz,\label{DSc}\\
\frac{\partial\,\Dz }{\partial\xi} &=& -\Dz\,\left[1+3\Dz\right]+\frac{\Omm}{2}\,\left[\Dz-\Dm\right]-\Oml\,\Dz.\label{DSd}
\ea\ese
which can be integrated by specifying initial conditions at a suitable initial hypersurface marked by constant $\xi$. Notice that the hypersurfaces of constant $\xi$ do not coincide with the $\T$ marked by constant $\tau\ne 0$ (though the initial hypersurface $\tau=0$ coincides with $\xi=0$). Hence, the radial gradient of a scalar $A$ along slices with constant $\tau \ne 0$ are different from those along slices with constant $\xi \ne 0$:
\begin{equation}\fl A'=\left[\frac{\partial A}{\partial r}\right]_\tau=\left[\frac{\partial A}{\partial r}\right]_\xi+\frac{\partial \xi}{\partial r}\,\left[\frac{\partial A}{\partial \xi}\right]_r=\left[\frac{\partial A}{\partial r}\right]_\xi+\frac{R_i'\,(\Gamma-1)}{R_i\,z_q}\,\left[\frac{\partial A}{\partial \tau}\right]_r,\label{difopr}\end{equation}
where we used (\ref{xidef}) and (\ref{Gamma}) to eliminate $\xi'$ and $\partial/\partial\xi$ in terms of $\Gamma$ and $\partial/\partial\tau$. Applying (\ref{difopr}) to (\ref{Omm})--(\ref{Oml}) leads to the gradients of $\Omm$ and $\Oml$ along surfaces $\xi =$ constant
\bse\ba \fl \left[\frac{\partial \Omm}{\partial r}\right]_\xi=\frac{3R_i'}{R_i}\Omm\,\Gamma\,\left[\Dm-2\Dz+\frac{\Gamma-1}{3\Gamma}\left(\Omm-2\Oml-1\right)\right],\label{Ommrxi}\\
\fl \left[\frac{\partial \Oml}{\partial r}\right]_\xi=-\frac{6R_i'}{R_i}\Oml\,\Gamma\,\left[\Dz-\frac{\Gamma-1}{3\Gamma}\left(1+\frac{\Omm}{2}-\Oml\right)\right],\label{Omlrxi}\ea\ese
which provide consistency conditions for (\ref{DSa})--(\ref{DSd}), as the mixed derivatives $\partial^2 /\partial\xi\partial r$ obtained from these equations coincide with those obtained from the system.

Equations (\ref{DSa})--(\ref{DSd}) are autonomous evolution PDE's that contain only derivatives with respect to $\xi$, and so the radial dependence of the functions enters as a parameter. Still, we are dealing with PDE's, while the usual dynamical systems techniques (as discussed in \cite{DS1,DS2,DS3,DS4}) have been conceived for autonomous systems of ODE's. However, as shown rigorously in Appendix B of \cite{suss08}, a system of PDE's like (\ref{DSa})--(\ref{DSd}) can be considered as a proper dynamical system, but one that is constrained by the fulfillment of the the initial conditions given explicitly at $\xi=0$ 
\bse\ba \Ommi = \frac{2\mu_{qi}}{2\mu_{qi}+\lambda-\kappa_{qi}},\label{CIa}\\
\Omli = \frac{\lambda}{2\mu_{qi}+\lambda-\kappa_{qi}},\label{CIb}\\
\Dim = \frac{R_i}{3R_i'}\left(\frac{[\Ommi]'}{\Ommi}-\frac{[\Omli]'}{\Omli}\right),\label{CIc}\\
\Diz = -\frac{R_i}{6R_i'}\frac{[\Omli]'}{\Omli},\label{CId}\ea\ese
where we used (\ref{Da_def}) and (\ref{Omm})--(\ref{Oml}) evaluated at $\xi=0$ (so that $L=\Gamma=1$). Once these conditions are satisfied in the initial hypersurface $\xi=0$ (which coincides with $\tau=0$), then we can apply to the numerical solutions of (\ref{DSa})--(\ref{DSd}) the standard techniques of dynamical systems. 
\begin{table}
\caption{Critical points of various invariant subsets. The letters {\bf{PA}} and {\bf{FA}} stand for past and future attractors (source/sink), while {\bf{SP}} is a saddle point. See (\ref{PA})--(\ref{SP5}) for reference of these critical points (except those of the Homogeneous subspace).}
\begin{tabular} {|l|lllll|}

    \hline

    Subspace & Critical &points & &&\\&&&&& \\ \hline

    Homogeneous    &
    $\Omega^{(\mu)} =1$,& $\Omega^{(\lambda)}=0$,& $\Dm=0$&$\Dz=0$,
    &{\bf{hPA}}\\
    subspace (FLRW)&$\Omega^{(\mu)}=0$,&$\Omega^{(\lambda)}=1$,&$\Dm=0$&$\Dz=0$,&{\bf{hFA}}\\
    $\Dm=\Dz=0$&$\Omega^{(\mu)} =0$,&$\Omega^{(\lambda)}=0$,&$\Dm=0$&$\Dz=0$,&{\bf{hSP2}}\\ \hline

    Spatially flat   & $\Omm
    =1$,&$\Oml=0$,&$\Dm=-1$,&$\Dz=-1/2$,&{\bf{PA}}\\
    subspace&$\Omm
    =0$,&$\Oml=1$,&$\Dm\,\hbox{arbitrary}$,&$\Dz=0$,&{\bf{FA}}\\
    $\Omm+\Oml=1$&$\Omm =1$,&$ \Oml=0$,&$ \Dm=-1$,&$\Dz=1/3$,&{\bf{SP3}}\\
                 &$\Omm =1$,&$ \Oml=0$,&$\Dm=0$,&$\Dz=0$,&{\bf{SP4}}\\
                 &$\Omm =0$,&$ \Oml=1$,&$\Dm=-1$,&$\Dz=-2/3$,&{\bf{SP5}} \\ \hline

    Schwarzschild-  & $\Omm =1$,&$
    \Oml=0$,&$\Dm=-1$,&$\Dz=-1/2$,&{\bf{PA}}\\
    deSitter subspace&$\Omm =0$,&$ \Oml=1$,&$\Dm=-1$,&$\Dz=0$,&{\bf{FA}}\\
    $\Dm=-1$&$\Omm=0$,&$\Oml=0$,&$\Dm =- 1$,&$\Dz=-1/3$,&\hbox{\bf{SP1}}\\
    &$\Omm=0$,&$\Oml=0$,&$\Dm=-1 $,&$\Dz  = 0$,&\hbox{\bf{SP2}}\\
    &$\Omm=
    1$,&$\Oml=0$,&$\Dm  =  - 1$,&$\Dz
    =1/3$,&\hbox{\bf{SP3}}\\
    &$\Omm =0$,&$\Omm=1$,&$\Dm=-1$,&$\Dz = -2/3$, &\hbox{\bf{SP5}}\\ \hline

    LTB dust & $\Omm=1$,&$\Oml=0$,&$\Dm=-1$,&$\Dz=-1/2$,&{\bf{PA}}\\
    without $\Lambda$&$\Omm =0$,&$\Oml=0$,&$\Dm\,\hbox{arbitrary}$,&$\Dz=0$,&{\bf{FA}}\\
    $\Oml=0$&$\Omm =0$,&$ \Oml=0$,&$\Dm=-1$,&$\Dz=-1/3$, &{\bf{SP1}}\\
    &$\Omm=1$,&$\Oml=0$,&$ \Dm=-1$,&$\Dz=1/3$,&\hbox{\bf{SP3}}\\
    &$\Omm=1$,&$\Oml=0$,&$\Dm=0$,&$\Dz  = 0$ ,&\hbox{\bf{SP4}} \\ \hline
\end{tabular}
\label{tab1}
\end{table}

\section{The phase space, invariant subsets and critical points.}

\subsection{The phase space in general.}

The solution curves $[\Omm(\xi,r),\,\Oml(\xi,r),\, \Dm(\xi,r),\, \Dz(\xi,r)]$ of the dynamical system (\ref{DSa})--(\ref{DSd}), obtained by numerical integration for initial conditions (\ref{CIa})--(\ref{CId}), evolve in a 4--dimensional phase space parametrized as
\begin{equation} \PP=\{\Omm,\,\Oml,\, \Dm,\, \Dz\}\subset\mathbb{R}^4.\label{Pspace}\end{equation}
and characterized  by the following critical points:

\begin{itemize}

\item {\underline{Source (past attractor):}}\,\,{\bf{PA}}
\begin{equation}\fl\Omm =1,\quad \Oml=0,\quad\Dm=-1,\quad\Dz=-1/2,\label{PA}\end{equation}
\item {\underline{Sink (future attractor):}}\,\,{\bf{FA}}
\begin{equation}\fl\Omm =0,\quad \Oml=1,\quad\Dm\,\,\hbox{arbitrary},\label{FA}\quad\Dz=0,\end{equation}
\item {\underline{Saddle Points:}}\,\,{\bf{SP}}
\ba   \fl\Omm=\Oml=0,\quad\left\{ \begin{array}{l}
 \Dm  =  - 1,\quad\quad\quad \Dz  = -1/3,\qquad\qquad\hbox{\bf{SP1}} \\
 \Dm\,\,\hbox{arbitrary} ,\qquad \Dz  = 0,\qquad\qquad\hbox{\bf{SP2}} \\
 \end{array} \right.\label{SP12}\ea
\ba   \fl\Omm=1,\quad\Oml=0,\quad\left\{ \begin{array}{l}
 \Dm  =  - 1,\qquad\; \Dz  = 1/3,\quad\;\;\hbox{\bf{SP3}} \\
 \Dm=0 ,\qquad\quad \Dz  = 0,\qquad\quad\;\hbox{\bf{SP4}} \\
 \end{array} \right.\label{SP34}\\
\fl \Omm =0,\quad \Oml=1,\quad \Dm=-1,\quad \Dz =
-2/3.\qquad\qquad \hbox{\bf{SP5}}.\label{SP5}\ea

\end{itemize}
As we show in the following subsections, these critical points are
common to various invariant subspaces. Notice that the future
attractor is a set of sinks defining a curve in $\PP$ parametrized
by $\Dm$ (the same remark applies to the saddle and the saddle {\bf{SP2}}).

\subsection{The homogeneous subspace and symmetry centers: FLRW dynamics.}

Bearing in mind (\ref{slaw_Dmk}), (\ref{slaw_Dh}) and (\ref{Gamma}), we have:  
\begin{equation}\Dim=\Diz=0 \qquad \Rightarrow\qquad \Dm=\Dz=0\quad \forall\;\; \xi\ne 0,\label{IC_HS}\end{equation}
which defines the ``homogeneous'' invariant subspace:
\begin{equation}\textrm{H}\equiv\{\Omm,\,\Oml,\,0,\,0\}\subset \PP,\label{Hsubset}\end{equation}
characterized by the 2--dimensional dynamical system (\ref{DSa})--(\ref{DSd}), which only involves $\Omm$ and $\Oml$, and is formally identical to the dynamical system associated with FLRW spacetimes with a dust plus $\Lambda$ source.  Analytic solutions for (\ref{DSa})--(\ref{DSd}) are given by (\ref{Omm}) and (\ref{Oml}) with  $L=\exp(\xi)$, and the critical points are listed in table \ref{tab1}: they are the projection of {\bf{PA}}, {\bf{FA}} and {\bf{SP2}}  on (\ref{Hsubset}).  Notice that, in general and irrespectively of initial conditions, the following limits hold for all curves in $\textrm{H}$:
\bse\ba \xi\to\infty\quad (\hbox{or}\quad L\to\infty)\quad \hbox{then}\quad \Omm\to 0,\,\,\Oml\to 1,\label{limit1}\\
\xi\to-\infty\quad (\hbox{or}\quad L\to 0)\quad \hbox{then}\quad \Omm\to 1,\,\,\Oml\to 0,\label{limit2} \ea\ese
Hence, the past and future attractors of  $\textrm{H}$ clearly correspond to the past and future limits $\xi\to\pm\infty$ of the inextensible trajectories in given by (\ref{Omm})--(\ref{Oml}).  

If $r=0$ marks a symmetry center, then (\ref{IC_HS}) must hold for all $\xi$ at $r=0$.  Hence, irrespectively of the initial conditions for $r> 0$, if a center worldline exists it is the only trajectory in $\PP$ whose phase space evolution takes place entirely in (\ref{Hsubset}). This is also valid if there is a second symmetry center marked by $r=r_c>0$.

\subsection{The spatially flat subspace.}

Bearing in mind the forms of $\Omm$ and $\Oml$ given by (\ref{Omm})--(\ref{Oml}),
we obtain the following important result
\begin{equation}\Ommi+\Omli-1 = 0 \quad \Rightarrow\quad \Omm+\Oml-1 = 0,\quad \forall\,\xi\ne 0.\label{sflat2}\end{equation}
Since $\Omm+\Oml=1$ is equivalent to $\kappa_q=0$, and $\kappa_q=0$ implies $\RR_q=E=0$, which (from (\ref{Theta1RR1})) implies $\RR=0$, the constraint (\ref{sflat2}) defines the spatially flat invariant subset
\begin{equation}\textrm{SF}\equiv\{\Omm,\,\Oml,\,\Dm,\,\Dz\,\,|\,\,\Omm+\Oml-1 = 0\}\subset \PP,\label{SF}\end{equation}
Notice that the intersection of $\textrm{SF}$ and $\textrm{H}$ is the straight line $\{\Omm(\xi)+\Oml(\xi)= 1\}$ in the $[\Omm,\Oml]$ plane. This is the invariant subspace that corresponds to the ``$\Lambda$--CDM'' model: a FLRW cosmology whose source is dust (CDM) and a cosmological constant. For $\Omm\sim 0.3$ and $\Oml\sim 0.7$, this model is often regarded as the best fit to cosmological observations at large scales and in the present cosmic era.

The 3--dimensional dynamical system associated with $\textrm{SF}$ follows from the general system by the restriction $\Oml=1-\Omm$. Its critical points are listed in table \ref{tab1}.  Notice that the past and future attractors exactly coincide with those of the general case. This indicates that all models have an initial asymptotic state with zero spatial curvature, while all models in which $\xi$ can be infinitely extended evolve towards a final zero curvature asymptotic state. The saddle points {\bf{SP1}} and {\bf{SP2}} are not critical points.

\subsection{The Kottler solution: vacuum Schwarzschild--deSitter subspace}

Consider initial conditions with arbitrary $\Ommi,\,\Omli,\,\Diz$ but $\Dim=-1$. Then, if there are no shell crossings ($\Gamma>0$) the scaling law (\ref{slaw_Dmk}) implies that $\Dm(\xi)=-1$ holds for all $\xi$. By inserting $\Dm=-1$ in (\ref{qvars}) and considering (\ref{rho_dust}), this constraint defines an invariant 3--dimensional subspace
\begin{equation} \textrm{V}\equiv\{\Omm,\,\Oml,\,-1,\,\Dz\}\subset \PP,\label{Vsubsp}\end{equation}
characterized by:  $\rho=0$, with $(\kappa_0c^2/3)\rho_q\,R^3=2\mu_q h_s^2 R^3=2M_0$ and $\mu_q=M_0/R^3$. These parameters identify the vacuum Kottler solution \cite{kottler} (or Schwarzschild--de Sitter) as the vacuum limit of $\Lambda$--LTB models with Schwarzschild mass $M_0$, and given in comoving coordinates constructed with radial geodesic observers (notice that (\ref{fried_dust}) with $M=M_0$ is the equation of these geodesics). 

The 3--dimensional dynamical system in this case follows by substituting (\ref{Vsubsp}) into (\ref{DSa})--(\ref{DSd}), while the corresponding critical points are listed in Table \ref{tab1}. 
Notice that the source {\bf{PA}} and all saddles except {\bf{SP4}} exactly coincide with those of the general case, whereas the future attractor {\bf{FA}} is contained in the future attractor of the general case: it is the point $\Dm=-1$ in the curve of sinks parametrized by $\Dm$.

\subsection{LTB dust without $\Lambda$.}

If initial conditions are selected with $\lambda=0=\Omli=0$, then $\Oml(\xi)=0$ holds for all $\xi$. In this case we have the invariant 3--dimensional subspace
\begin{equation} \textrm{\bf{D}}\equiv\{\Omm,\,0,\,\Dm,\,\Dz\}\subset \PP,\label{Dsubsp}\end{equation}
associated with the standard dust LTB solutions without cosmological constant. This subset was examined in detail in \cite{suss08}
\footnote{The phase space variables $\Omm,\,\Dm$ and $\Dz$ were denoted in \cite{suss08} by $\langle\Omega\rangle,\,\Delta^{(m)}$ and $-S$.}.
The corresponding 3--dimensional dynamical system follows by setting $\Oml=0$ in (\ref{DSa})--(\ref{DSd}), with the critical points listed in Table \ref{tab1}.  Notice that the source exactly coincides with the source {\bf{PA}} of the general case, indicating how the cosmological constant plays no role in the early evolution (near the initial curvature singularity $\xi\to -\infty$). The saddle points {\bf{SP2}} and {\bf{SP5}} are not critical points of this subsystem. The future attractor {\bf{FA}} is different from that of the general case, which indicates that $\lambda$ plays a distinctive dominant role in the late evolution in layers in which $\xi\to\infty$.

\section{Kinematics of dust layers.}

\begin{figure}[htbp]
\begin{center}
\includegraphics[width=4in]{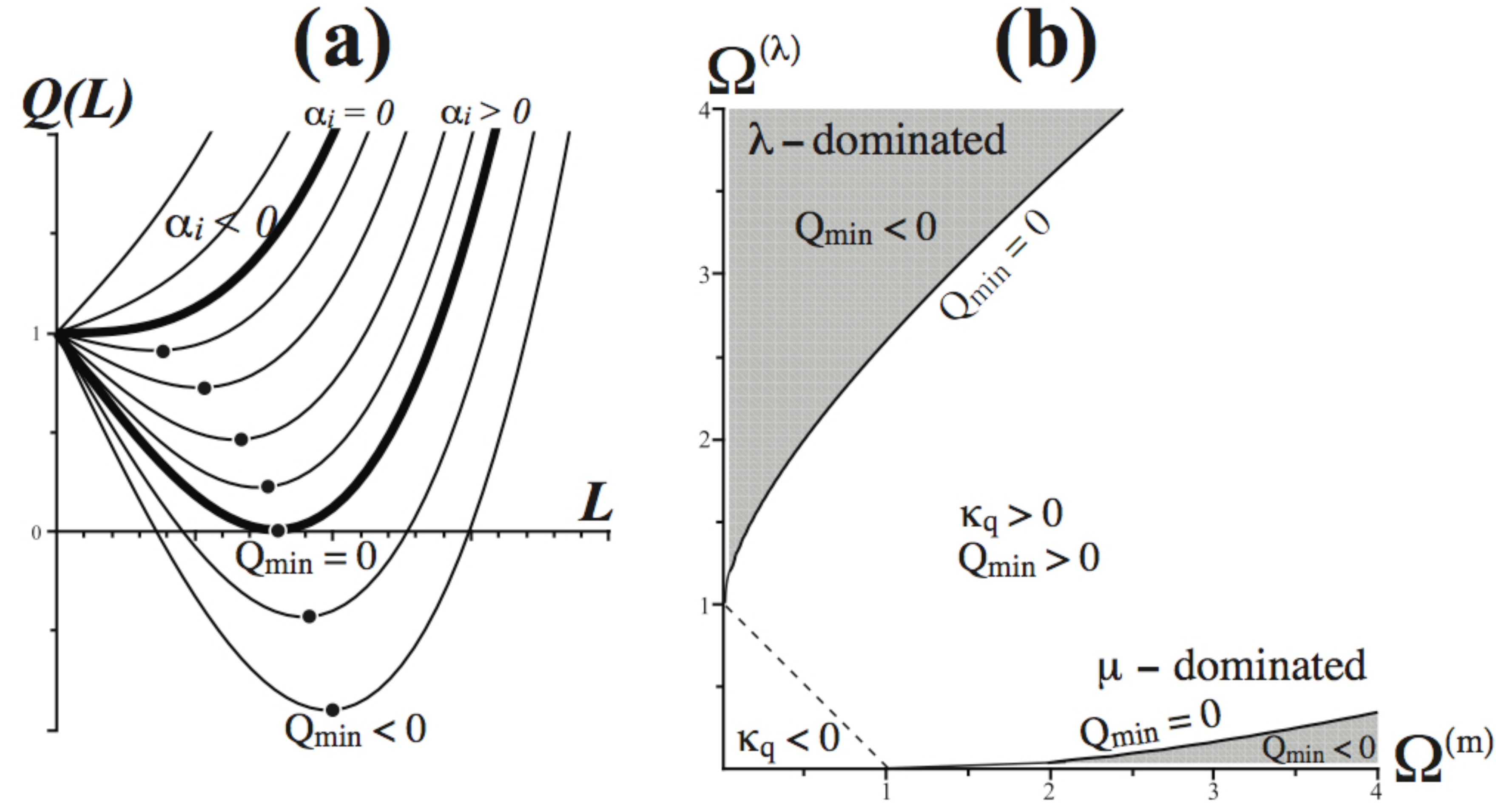}
\caption{{\bf Initial conditions and Kinematics.} Panel (a) shows the cubic polynomial $Q(L)$ defined in (\ref{cubeq}), whose roots determine the zeroes of $L_{\tau}$. Configurations corresponding to $\Qmin>0$ (between the two thick curves) are perpetually expanding (no zeroes) but have positive spatial curvature. Configurations with $\Qmin\leq 0$ must bounce or collapse. Panel (b) shows the regions in the space of initial conditions $\{\Ommi,\Omli\}$ for which $\Qmin<0$ holds (shaded areas).}
\label{fig1}
\end{center}
\end{figure}

The relation between the possible type of kinematic evolution and initial conditions can be appreciated by looking at the roots of $L_{,\tau}$ in equation (\ref{sqHq}), or equivalently, by  looking at the positive real roots of the cubic $Q(L)$, which we rewrite as
\bse\ba Q(L) = \beta_i\, L^3-\alpha_i\, L + 1 = 0,\label{cubeq}\\
\beta_i = \frac{\Omli}{\Ommi}=\frac{\lambda}{2\mu_{qi}},\qquad \alpha_i = \frac{\Ommi+\Omli-1}{\Ommi}=\frac{\kappa_{qi}}{2\mu_{qi}}.\ea\ese
While (\ref{cubeq}) can be solved exactly, it is more practical to examine it by looking at the qualitative behavior of the cubic polynomial $Q(L)$ (see figure \ref{fig1}a). The following cases arise
\begin{itemize}
\item If $\alpha_i\leq 0$ then $Q(L)$ has no extrema and $Q(L)\geq 1$ holds for all $L$. This case clearly corresponds to the perpetually expanding/collapsing configuration with $\Ommi+\Omli<1$ (or  $\kappa_{qi}\leq 0$).
\item If $\alpha_i> 0$ then $Q(L)$ has a minimum at $L=\Lmin=[\beta_i/(3\alpha_i)]^{1/2}$, so that
\begin{equation}\fl \Qmin=Q(\Lmin)=1-\frac{\kappa_{qi}^{3/2}}{3\sqrt{3\lambda}\,\mu_{qi}}=1-\frac{2\left[\Ommi+\Omli-1\right]^{3/2}}{3\sqrt{3}\,\left[\Omli\right]^{1/2}\Ommi}.\label{Qmindef}\end{equation}
Clearly, since $Q(L)\geq \Qmin$, then (\ref{cubeq}) has no positive roots if $\Qmin>0$ (see figure \ref{fig1}a), which corresponds to perpetually expanding/collapsing configurations with $\Ommi+\Omli>1$ (or $\kappa_{qi}> 0$).  However, a sufficient condition for the existence of positive roots of (\ref{cubeq}) is
\begin{equation} \Qmin(\Ommi,\Omli)\leq 0,\label{Qmin}\end{equation}
which corresponds to the configurations that have turning values in which the range of $L$ is restricted by a bounce where $L_{,\tau}=0$.

\end{itemize}
In order to examine the initial conditions associated with the possible kinematic evolutions described above, we examine the constraint (\ref{Qmin}) by plotting in figure \ref{fig1}b the level curves of $Q_{\textrm{\tiny{min}}}$ in the parameter plane $\{\Ommi,\Omli\}$. Since these initial value functions depend on $r$, then any given set of initial conditions will appear as a curve $[\Ommi(r),\Omli(r)]$ in this plane. The regions where (\ref{Qmin}) holds are marked by the gray areas in the upper left and lower right corners in figure \ref{fig1}b, respectively corresponding to a ``$\lambda$--dominated'' and ``$\mu$--dominated'' regimes. It is interesting to notice that the ``$\mu$--dominated'' regime is more restrictive for the fulfillment of (\ref{Qmin}), since a larger $\Ommi$ is needed even for small $\Omli<1$.

The kinematic evolution patterns of dust layers in $\Lambda$--LTB models and their relation with initial conditions are described below:

\begin{enumerate}

\item {\underline {Perpetual expansion (collapse) from (to) a singularity}}. If $\kappa_{qi}\leq 0$, or if $\kappa_{qi}>0$ but $\Ommi$ and $\Omli$ are roughly comparable for all $r$, we are well within the non--shaded area of figure \ref{fig1}b and (\ref{Qmin}) does not hold.  See figures \ref{fig4}, \ref{fig5} and \ref{fig6}.

\item {\underline {All layers initially expand from and then collapse to a singularity}}. This pattern corresponds to the extreme ``$\mu$--dominated'' regime in figure \ref{fig1}b and to the range restriction $0<L<L_1$ with the bounce $L_{,\tau}=0$ at $L=L_1$. Initial conditions must comply with $\Ommi\gg 1$ and $\Omli\ll 1$ for all $r$. If $\Oml=0$ (pure dust) then $\kappa_{qi}>0$ is a sufficient condition for this evolution, but $\kappa_{qi}>0$ is only necessary (not sufficient) if  $\Oml>0$. See figure \ref{fig7}.

\item {\underline {All layers initially collapse from infinity, bounce and re--expand}}. This pattern corresponds to the ``$\lambda$--dominated'' regime in figure \ref{fig1}b, and to the  range restriction $L>L_2$ where the bounce of $L$ occurs at $L=L_2$, so that layers avoid the collapse.  Initial conditions must comply with $\kappa_{qi}>0$,\, $\Ommi\ll 1$ and $\Omli\gg 1$ holding for all layers. See figure \ref{fig8}.

\item {\underline {Loitering evolution}}. Layers initially expanding or collapsing reach an unstable static state. Layers may collapse or expand at later times ($\Lambda$--dust FLRW models with this kinematics are known as Eddington--Lema\^\i tre models). Initial conditions associated with this pattern comply with $\kappa_{qi}>0$ and $\Qmin= 0$. See figure \ref{fig9}.

\item {\underline {Mixed evolution patterns}}. Since $\Ommi,\Omli$ depend on $r$, dust layers may show different evolution patterns in different radial ranges (as opposed to the homogeneous FLRW case). In these cases initial conditions must be specified so that (\ref{Qmin}) only holds in part of the radial range, or equivalently, with the curve $[\Ommi(r),\Omli(r)]$ passing from the non--shaded area to either one of the shaded areas of figure \ref{fig1}b. Two possible configurations are:

\begin{itemize}
 \item Layers around the center follow pattern (ii) and ``external'' layers evolve as in (i). See figure \ref{fig10}.
 \item Layers around the center follow pattern (i) and  ``external'' layers evolve as in (iii). See figure \ref{fig11}.
\end{itemize}

\end{enumerate}
We examine the phase spaces of configurations following the evolution patterns (i)--(v) in  section 9. See \cite{LCDM} for detailed plots of the scale factor for patterns (i)--(iv) in FLRW cosmologies.  

\section{Topology of the space slices and a choice of radial coordinate.}

The LTB metric (in both forms (\ref{LTB}) and (\ref{LTB2})) admits an arbitrary rescaling of the radial coordinate. We use this coordinate gauge freedom to fix the function $R_i$ in (\ref{LGdef}) and (\ref{LTB2}) by the choice
\begin{equation} R_i = h_s^{-1}\,f(r),\label{eqRi}\end{equation}
where $f$ is a smooth dimensionless function, while $h_s^{-1}$ (defined in (\ref{dimless})) provides a length scale for the radial coordinate. The topological class of the slices $\T[t]$ is given by the number of symmetry centers, which determines the allowed forms of $f$. Assuming absence of shell crossings, so that $\Gamma>0$ holds everywhere, the admissible topologies of the slices $\T[t]$ are

\begin{itemize}
\item {\underline{``Open models''}}. The $\T[t]$ are homeomorphic to $\mathbb{R}^3$ and admit only one symmetry center at $r=0$. Since $R'>0$ and $R'_i>0$ must hold for all $r$, then $f$ can be any monotonously increasing function complying with $f(0)=0$. The simplest choice is $f(r)=r$, but we will use instead $f(r)=\tan r$ so that $R_i\to\infty$ corresponds to the finite limit $r\to\pi/2$.
\item {\underline{``Closed models''}}. The $\T[t]$ are homeomorphic to $\mathbb{S}^3$ and
admit two symmetry centers, at $r=0$ and $r=r_c$. Since $R(t,0)=R(t,r_c)=0$ for all $t$, then there must exist (at least) a turning value $r=r_{\textrm{\tiny{tv}}}$ so that $R'(ct,r_{\textrm{\tiny{tv}}})=0$ where $0<r_{\textrm{\tiny{tv}}}<r_c$. The function $f$ must comply with $f(0)=f(r_c)=0$ and $f'(r_{\textrm{\tiny{tv}}})=0$. A convenient choice is $f(r) =\sin r$.
\item {\underline{``Wormholes''}}. The $\T[t]$ can be homeomorphic to $\mathbb{S}^2\times \mathbb{R}$ or $\mathbb{S}^2\times \mathbb{S}^1$. While the $\T[t]$ do not admit symmetry centers, there must be (at least) a turning value $R'(ct,r_{\textrm{\tiny{tv}}})=0$. The function $f$ cannot have zeroes, but must comply with $f'(r_{\textrm{\tiny{tv}}})=0$. We will use $f(r)=\sec r$ in the examples of section 9.
\end{itemize}
In order to examine the relation between the topological class of the $\T[t]$, the sign of the spatial curvature and the kinematic evolution of dust layers, we remark that, even if  $\Gamma>0$ holds, $R_i$ must comply with the following condition associated with the regularity of the $g_{rr}$ metric coefficient in (\ref{LTB}) and (\ref{LTB2})\cite{sussLTB1,sussLTB2,ltbstuff,suss02}
\begin{equation} \frac{R'_i}{\left| 1-\kappa_{qi}R_i^2 h_s^2\right|^{1/2}}=\frac{h_s^{-1}\,f'}{\left| 1-\kappa_{qi}f^2\right|^{1/2}}\ne 0.\label{grr1}\end{equation}
Hence, if there is a turning value $r=r_{\textrm{\tiny{tv}}}$ such that $R'_i(r_{\textrm{\tiny{tv}}})=0$, the denominators in the expressions above must have a same order zero at $r=\rtv$. This places the following restriction on $f$ and $\kappa_{qi}$
\begin{equation} f'(\rtv)=0\quad \Leftrightarrow\quad  \kappa_{qi}(r_{\textrm{\tiny{tv}}})=\frac{1}{f^2(r_{\textrm{\tiny{tv}}})}>0.\label{grr2}\end{equation}
If (\ref{grr1})--(\ref{grr2}) fail to hold we have a surface layer singularity at $r=r_{\textrm{\tiny{tv}}}$ \cite{sussLTB1,ltbstuff,suss02}. Therefore, (\ref{grr2}) implies that all regular models with closed and wormhole topologies must have positive spatial curvature: $\kappa_{qi}>0$ or $\Ommi+\Omli-1>0$. In the case of pure dust LTB models ($\lambda=0$), the implication of (\ref{grr2}) is that dust layers of all models with these topologies must bounce and collapse (as $\kappa_{qi}>0$ implies that $L_{,\tau}$ has necessarily a zero). However, the situation is different if $\lambda>0$, since dust layers can expand for all $\tau$ when $\kappa_{qi}>0$ as long as the initial value functions are selected so that (\ref{Qmin}) does not hold (see figure \ref{fig1}). Hence, fully regular and perpetually expanding  LTB--$\Lambda$ models with closed or wormhole topologies are possible.

\section{Phase space dynamics in terms of homogeneous and inhomogeneous projections.}

Since the homogeneous invariant subspace $\textrm{H}$ defined by (\ref{Hsubset}) corresponds to dust--$\Lambda$ FLRW cosmologies, the dynamical variables $\Omm$ and $\Oml$ of $\textrm{H}$ become the standard FLRW Omega parameters, and the dynamical system for $\textrm{H}$ (which is (\ref{DSa}) and (\ref{DSb})) is equivalent to the dynamical system of these FLRW cosmologies (in the expansion normalized variables \cite{DS1,DS2,DS3,DS4}). We define for the latter the 2--dimensional phase space
\begin{equation}\textrm{H}_{(0)}=\{\Omega^{(\mu)}, \Omega^{(\lambda)}\} \subset \mathbb{R}^2,\qquad \hbox{such that}\qquad \textrm{H} = i(\textrm{H}_{(0)})\label{H0}\end{equation}
where $i$ is the inclusion map $i: \mathbb{R}^2\to \mathbb{R}^4$,\,\, $i(x^1,x^2)=[x^1,x^2,0,0]$.

Since $\textrm{H}_{(0)}$ is endowed  with a clear physical and geometric meaning (the phase space of the homogeneous limit FLRW spacetime), while $\{\Dm,\Dz\}$ convey the deviation from homogeneity in the dynamics (and are gauge invariant covariant non--linear perturbations \cite{sussPRD}), it is useful to express the general phase space $\PP$ as the direct product
\begin{equation} \PP = \textrm{H}_{(0)} \times \textrm{I},\qquad \textrm{I}\equiv \{\Dm,\,\Dz\}\subset \mathbb{R}^2,\label{dirprod}\end{equation}
so that trajectories and critical points of $\PP$ can be described and analyzed by means of the following projection maps applied to every $\textrm{p}=[\Omm,\,\Oml,\,\Dm,\,\Dz]\in \PP$:
\bse\ba \fl \textrm{Homogeneous projection.}\nonumber\\
 \Pi_{\textrm{\tiny{H}}}:\PP\to \textrm{H}_{(0)},\quad\textrm{such that}\quad  \Pi_{\textrm{\tiny{H}}}(\textrm{p})=[\Omm,\,\Oml]\in\textrm{H}_{(0)},\label{Hproj}\\
\fl \textrm{Inhomogeneous projection.}\nonumber\\
 \Pi_{\textrm{\tiny{I}}}:\PP\to \textrm{I},\quad\textrm{such that}\quad  \Pi_{\textrm{\tiny{I}}}(\textrm{p})=[\Dm,\,\Dz]\in\textrm{I}.\label{Iproj}
\ea\ese
While the use of (\ref{Hproj}) and (\ref{Iproj}) solves the practical problem of dealing with the 4--dimensional phase space $\PP$, these projections highlight interesting dynamical features of $\PP$ which we discuss below.

Since the dynamical equations (\ref{DSa}) and (\ref{DSb}) associated with $\textrm{H}_{(0)}$ are completely independent of $\Dm,\,\Dz$ (the coordinates of $\textrm{I}$), there is a one--one equivalence map between every FLRW phase space trajectory in $\textrm{H}_{(0)}$ and a phase space trajectory of $\PP$ projected into $\textrm{H}_{(0)}$ by $\Pi_{\textrm{\tiny{H}}}$. However, the initial conditions for each trajectory in $\textrm{H}_{(0)}$ are two constants: $\Omega_i^{(\mu)},\,\Omega_i^{(\lambda)}$, and so each phase space trajectory corresponds to a single spacetime configuration. On the other hand, because of the inhomogeneity of $\Lambda$--LTB spacetimes, initial conditions are given (in general) by functions of $r$. Hence, these trajectories form a one--parameter family of curves, so that each curve (when projected into $\textrm{H}_{(0)}$ by $\Pi_{\textrm{\tiny{H}}}$) is equivalent to a trajectory of a distinct FLRW spacetime.

The critical points of $\textrm{H}_0$ are simply the critical points of $\textrm{H}$ (see table \ref{tab1}) projected under  $\Pi_{\textrm{\tiny{H}}}$, and correspond to the particular case spacetimes:
\bse\ba\Omm=1,\,\Oml=0\qquad \hbox{Past attractor: spatially flat FLRW},\label{PAH0}\\
  \Omm=0,\,\Oml=1\qquad \hbox{Future attractor: de Sitter},\label{FAH0}\\
   \Omm=0,\,\Oml=0\qquad \hbox{Saddle: Minkowski},\label{SH0}  \ea\ese
However, each one of the critical points in $\PP$ listed in section 5 is also projected by $\Pi_{\textrm{\tiny{H}}}$ into one of the critical points of $\textrm{H}_0$ listed above:
\begin{itemize}
\item {\bf{PA}}, {\bf{SP3}} and {\bf{SP4}} are projected into the past attractor (\ref{PAH0})
\item {\bf{FA}} and {\bf{SP5}}  are projected into the future attractor (\ref{FAH0})
\item {\bf{SP1}} and {\bf{SP2}}  are projected into the saddle (\ref{SH0})
\end{itemize}
These projections  clearly suggest characterizing every critical point in $\PP$ as related to a given limiting spacetime, such as spatially flat FLRW, de Sitter or Minkowski. The degeneracy in projecting the critical points of $\PP$ into any one of the critical points (\ref{PAH0})--(\ref{SH0}) occurs when we only look at the projection of the trajectories of $\PP$ into $\textrm{H}_{(0)}$ by $\Pi_{\textrm{\tiny{H}}}$, but it disappears when we examine the projection of these curves into $\textrm{I}$ by means of (\ref{Iproj}), whose coordinates $\{\Dm,\,\Dz\}$ convey the effects of inhomogeneity. Hence, a complete picture of the evolution of the trajectories in $\PP$ and their relation with the critical points follows when we combine the description that follows from the two projections (\ref{Hproj}) and (\ref{Iproj}).

The projections (\ref{Hproj}) and (\ref{Iproj}) will be used in the following section to analize the phase space trajectories of various representative $\Lambda$--LTB configurations associated with the kinematic patterns and topologies of the space slices discussed in sections 6 and 7.

\section{Numerical and graphical examples.}

We examine, by means of numeric solutions of  (\ref{DSa})--(\ref{DSd}) under initial conditions (\ref{CIa})--(\ref{CId}), the phase space evolution of various representative $\Lambda$--LTB configurations. By plotting the curve $[\Ommi(r),\Omli(r)]$ in the parameter space of figure \ref{fig1}b, we can infer for any given set of initial value functions a kinematic evolution pattern (i)--(iv) described in section 6. The topological class of the space slices follows from the choice of $f(r)$ in (\ref{eqRi}). To avoid dealing with shell crossing singularities, we verify numerically for every configuration under study the fulfillment of $\Gamma>0$ (notice that $\Gamma_{,\tau}=3\,\Gamma\,z_q\,\Dz$ follows from (\ref{Sigma1}), (\ref{SE}) and (\ref{Gamma})).

Since the phase space $\PP$ is 4--dimensional, we plot trajectories (curves parametrized by $\xi$ with fixed $r$) for any given set of initial conditions in eight figures composed by two panels, each one containing a 2--dimensional graph corresponding to:
\begin{itemize}
\item Panel ``{\bf (a)}'' (left hand side): the congruence of curves $[\Omm(\xi,r),\Oml(\xi,r)]$ given by the homogeneous projection of $\PP$ defined by (\ref{Hproj}).
\item Panel ``{\bf (b)}'' (right hand side): the congruence of curves $[\Dm(\xi,r),\Dz(\xi,r)]$ given by the inhomogeneous projection of $\PP$  defined by (\ref{Iproj}).
\end{itemize}
These graphs display the congruences of the trajectories, together with thick black dots marking critical points (past and future attractors, {\bf \textrm{PA}} and {\bf \textrm{FA}}, and saddle points {\bf \textrm{SP1}}, {\bf \textrm{SP2}}, etc). In configurations that contain dust layers that collapse or bounce (subsections 9.2 and 9.3, figures 6--9) the phase space variables $\Omm,\,\Oml$ and $\Dz$ diverge as $z_q\to 0$. For these cases we plot the $\arctan$ of the variables, so that $\pm$infinity is brought to $\pm\pi/2$. However, the trajectories cannot be continued by numerical integration of (\ref{DSa})--(\ref{DSd}) beyond these points. In these cases we must use the system (\ref{sys1a})--(\ref{sys1d}) with the variable $s=z_q\,\Dz$ (which is bounded as $z_q\to 0$) to examine the behavior of $\Omm,\,\Oml$ and $\Dz$ in the full evolution range.

\subsection{Perpetually expanding configurations: kinematic pattern (i).}

If initial conditions are chosen so that the curve $[\Ommi(r),\Omli(r)]$ lies entirely in the non--shaded region of figure \ref{fig1}b and $z_{qi}>0$ for all $r$, the resulting configurations expand or collapse monotonously for all $\xi$. Since a collapsing evolution would follow from an expanding one by a trivial sign inversion of $\xi$, we will only consider the latter in the examples below.

\subsubsection{Negative spacial curvature.}

\begin{figure}[htbp]
\begin{center}
\includegraphics[width=4.0in]{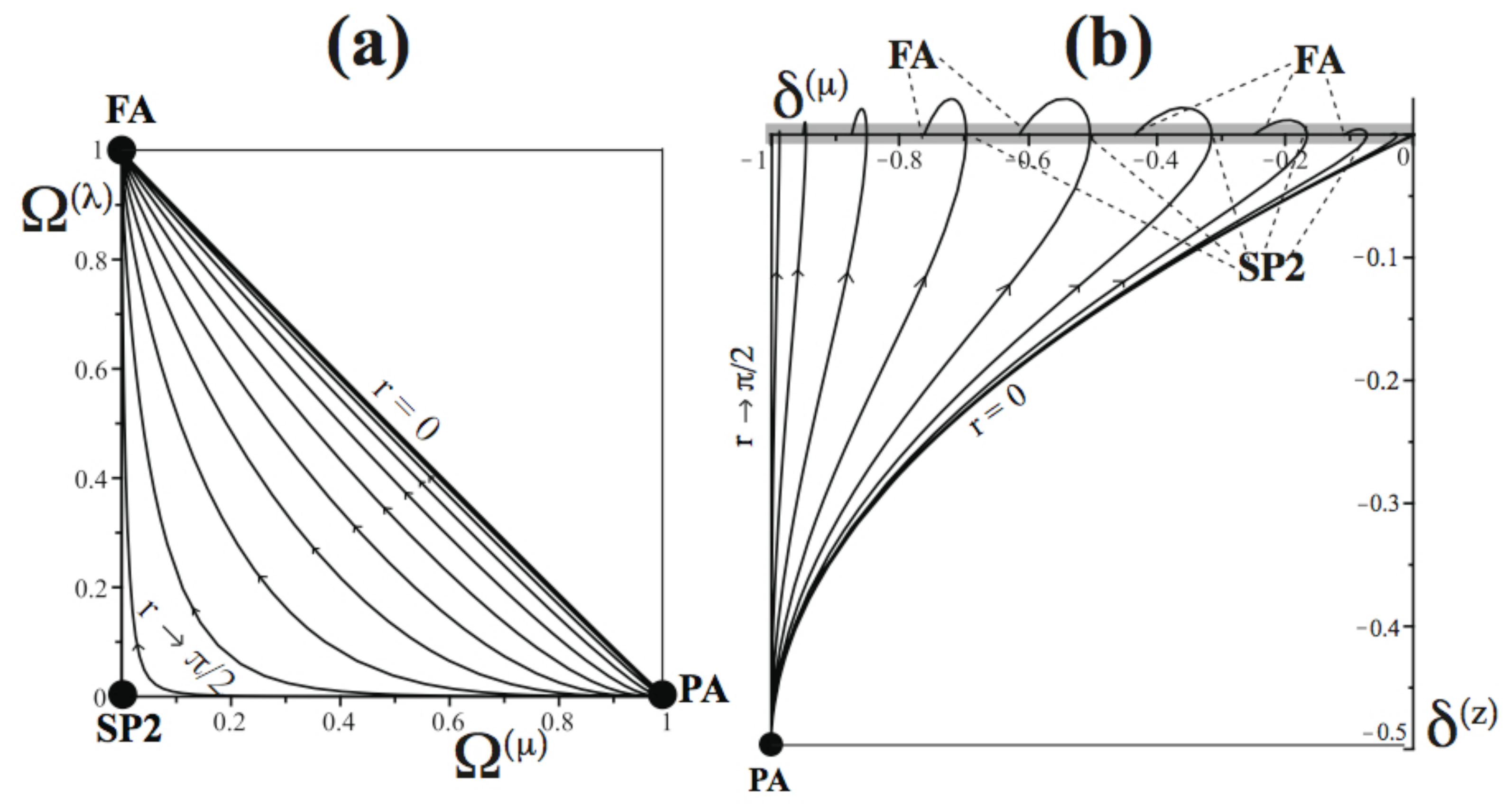}
\caption{{\bf Expanding configuration with negative curvature and asymptotic to Schwarzschild de Sitter.} These phase space trajectories follow from initial conditions (\ref{IVF_11a})--(\ref{IVF_11c}).}
\label{fig4}
\end{center}
\end{figure}

We consider $f(r) = \tan r$ in (\ref{eqRi}) in the range $0 \leq r < \pi/2$, so that $R_i\to\infty$ as $r\to\pi/2$. We consider first initial value functions given by
\bse\ba \fl \mu_{qi}=m_{10}+\frac{m_{11}-m_{10}}{1+f^3(r)},\quad &
(m_{10}=0, \quad m_{11}=15.3), &\label{IVF_11a}\\
\fl \kappa_{qi}=k_{10}+\frac{k_{11}-k_{10}}{1+
f^4(r)},&(k_{10}=-1.2,\quad k_{11}=-0.1),\quad &\label{IVF_11b}\\
\fl \lambda=0.1.&&\label{IVF_11c}\ea\ese
The phase space trajectories are displayed by figure \ref{fig4}, which highlights the past attractor {\bf \textrm{PA}} from which all trajectories emerge (and can be associated with the initial singularity). These curves approach the saddle point  {\bf \textrm{SP2}} and terminate in a point of the line of the future attractors {\bf \textrm{FA}} (thick gray line in panel (b)). because of the degeneracy in the $\Dm$ coordinate in {\bf \textrm{SP2}} and {\bf \textrm{FA}}, the curves in panel (b) cross the $\Dz$ axis at different values of $\Dm$. Since $\mu_{qi}\to 0$ as $R_i\to\infty$, this configuration must be radially asymptotic to a Schwarzschild de--Sitter vacuum. This is consistent with the fact that in the limit $r\to\pi/2$ (which implies $R_i\to\infty$) we have near vacuum conditions $\Oml\approx 0$ and $\Omm\approx 0$ (see \cite{sussLTB2}), and thus the trajectories in panel (a) tend to the lower left corner saddle in the intermediary stage of the evolution between {\bf \textrm{PA}} and {\bf \textrm{FA}}. Also, the curves in panel (b) in the limit $r\to\pi/2$ reach {\bf \textrm{FA}} with $\Dm\to -1$, which characterizes the invariant subspace of a Schwarzschild de--Sitter vacuum (see section 5.4).
\begin{figure}[htbp]
\begin{center}
\includegraphics[width=4.0in]{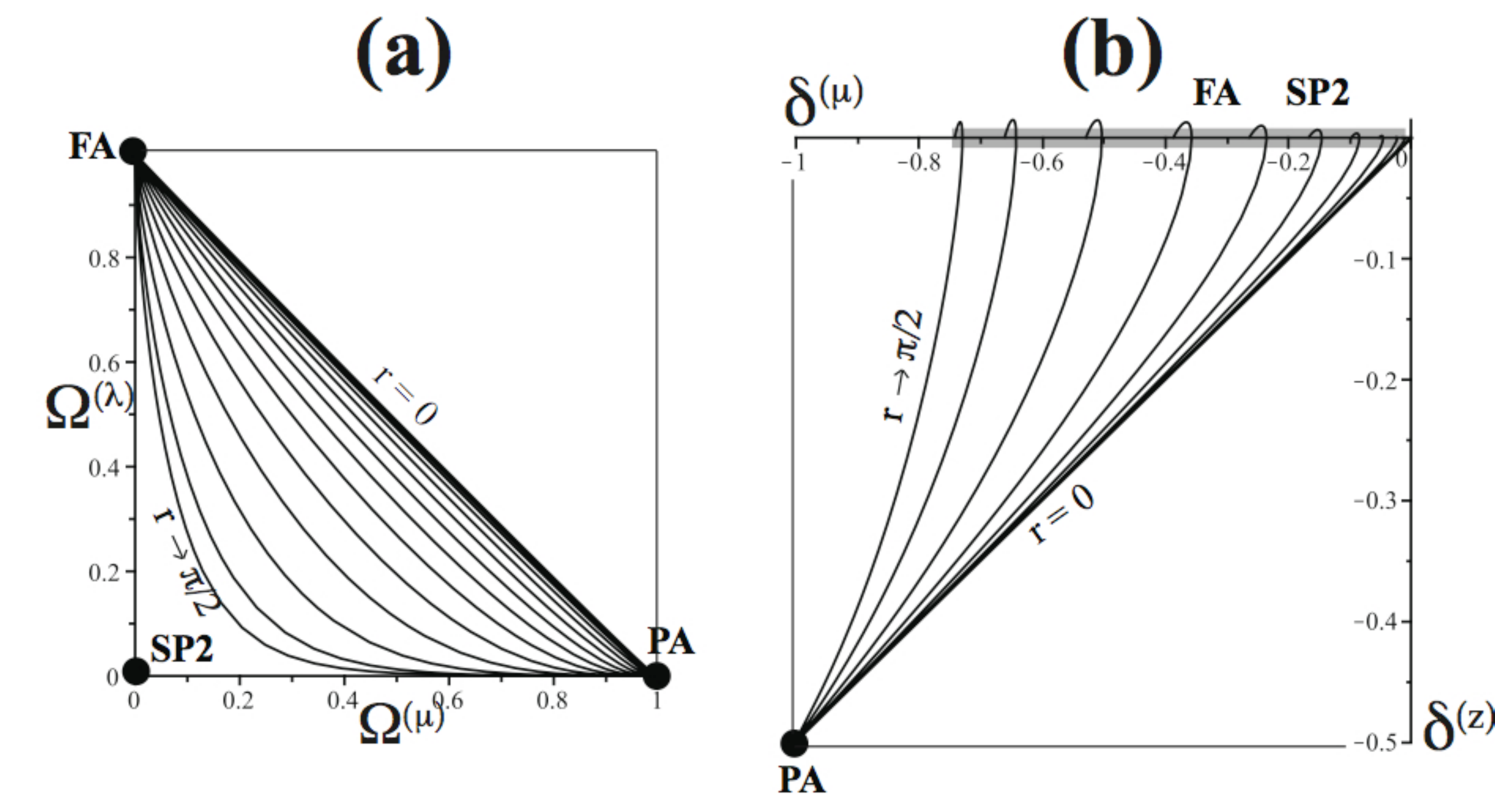}
\caption{{\bf Expanding configuration with negative curvature and asymptotic to FLRW.} Initial conditions and figure conventions are as those used in figure \ref{fig4}, but with $\mu_{qi}$ tending asymptotically to a constant.}
\label{fig5}
\end{center}
\end{figure}

If we consider initial conditions as in (\ref{IVF_11a})--(\ref{IVF_11c}), but with $m_{10}=0.1,\,k_{10}=-0.75$, so that $\mu_{qi}\to 0.1$ and $\kappa_{qi}\to -0.75$ as $R_i\to\infty$, we obtain a configuration with the same time evolution but radially asymptotic to a $\Lambda$--dust FLRW model with negative curvature \cite{sussLTB2}. The phase space trajectories (displayed by figure \ref{fig5}) are qualitatively analogous to those of figure \ref{fig4}, but the different radial asymptotic conditions imply that the trajectories as $r\to\pi/2$ ($R_i\to\infty$) do not reach as far as the lower left corner in panel (a) and do not reach as far as $\Dm\to-1$ in panel (b) (as expected if the model is radially asymptotic to a $\Lambda$--dust FLRW model).

\subsubsection{Positive spatial curvature with open topology.}

\begin{figure}[htbp]
\begin{center}
\includegraphics[width=4.0in]{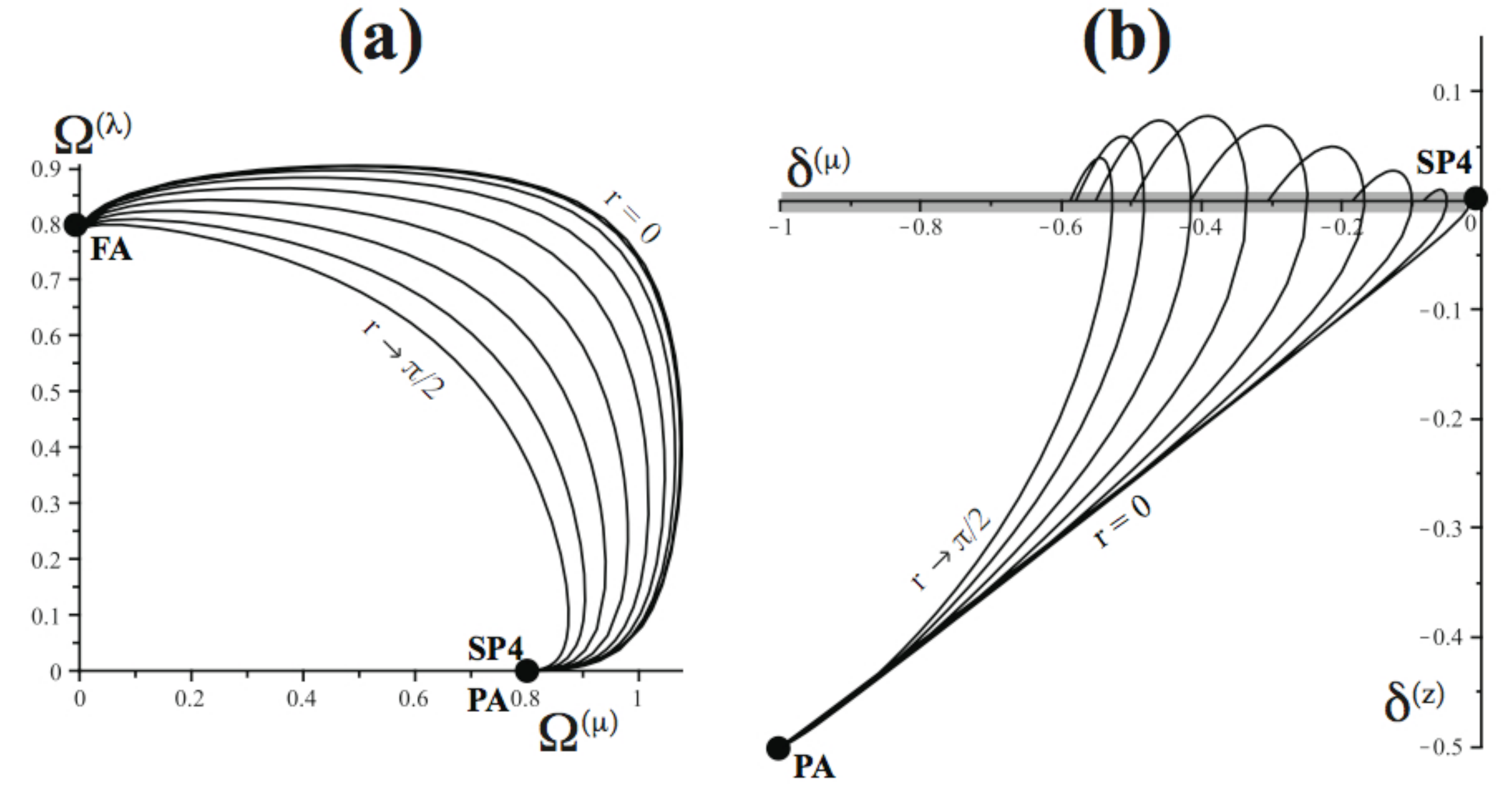}
\caption{{\bf Expanding open model with positive curvature and open topology.} Initial conditions are given by (\ref{IVF_21a})--(\ref{IVF_21c}).  See the text for further explanation.}
\label{fig6}
\end{center}
\end{figure}

We choose $f(r) = \tan r$ in the range $0\leq r<\pi/2$ for the following initial value functions
\bse\ba \fl \mu_{qi}=m_{10}+\frac{m_{11}-m_{10}}{1+f^2(r)},\quad &
(m_{10}=0.8, \quad m_{11}=30.0), &\label{IVF_21a}\\
\fl \kappa_{qi}=k_{10}+\frac{k_{11}-k_{10}}{1+
f^2(r)},&(k_{10}=0.7,\quad k_{11}=15.5),\quad &\label{IVF_21b}\\
\fl \lambda=0.75.&&\label{IVF_21c}\ea\ese
These configurations have no equivalent in pure dust models  $\Oml=\lambda=0$, as in this case positive curvature necessarily implies reversal of expansion and collapse (see section 6).  The phase space trajectories are displayed in figure \ref{fig6}. As in the previous cases the curves emerge from the past attractor {\bf \textrm{PA}} (big bang), expand monotonously and end in the future attractor {\bf \textrm{FA}}, but now the trajectories in panel (a) are located above the line $\Omm+\Oml=1$ characteristic of zero spatial curvature. The curves now approach the saddle point {\bf \textrm{SP4}}, whose projection coincides with the projection of {\bf \textrm{PA}} in the $\{\Omm,\Oml\}$ plane (panel (a)). 
\subsubsection{Positive spatial curvature with closed or wormhole topology.}

As in the previous case, these configurations have no equivalent in the pure dust case. 
\begin{itemize}
\item Closed topology. We take $f(r) = \sin r$, so that the radial range is restricted by $0\leq r\leq \pi$, with two symmetry centers at $r=0,\,\pi$, and $r=\pi/2$ marking the ``equator'' of the 3--sphere $\mathbb{S}^3$ where $R$ takes its maximal value and $R'(\pi/2)=0$ (see section 7).
\item Wormhole. We can use $f(r) = \sec r$, so that $-\pi/2< r < \pi/2$ and $R$ has no zeroes (no symmetry centers), while $R\to\infty$ as $r\to\pm\pi/2$. This choice yields space slices homeomorphic to $\mathbb{S}^2\times\mathbb{R}$, where the locus $r=0$ can be identified with the ``throat'' of the wormhole where $R$ takes its minimal value and $R'(0)=0$.
\end{itemize}
We do not display the phase space trajectories for these cases because for all choices of initial conditions they are very similar to those of the open models of figure \ref{fig6}, though some distinctive features are worth remarking. In the case of closed topology the trajectories of the two centers in the homogeneous projection (panel (a)) are identical to each other and to the curve for the center ($r=0$) in panel (a) of figure \ref{fig6}.
The trajectories in the range $\pi/2< r< \pi$ can be uniquely mapped to those in the range $0< r < \pi/2$. The curve for $r=\pi/2$ (the ``equator'' of $\mathbb{S}^3$) is analogous to the curve close to the line $\Omm+\Oml=1$ corresponding to radial infinity in the open models ($r\to\pi/2$ in both panels of figure \ref{fig6}). In the wormhole case the curves for $r<0$ and $r>0$ (``left'' and ``right'' of the throat $r=0$) are identical and can be mapped to each other, while the phase space curve of the throat is analogous to that of the center $r=0$ in both panels of figure \ref{fig6}. 


\subsection{Configurations in which expansion is reversed or halted for all dust layers.}

These configurations follow by choosing initial conditions complying with (\ref{Qmin}), so that the resulting curve $[\Ommi(r),\Omli(r)]$ is fully contained either in the ``$\mu$--dominated'' or the ``$\lambda$---dominated'' regimes (shaded areas in figure \ref{fig1}b).
\begin{figure}[htbp]
\begin{center}
\includegraphics[width=4.0in]{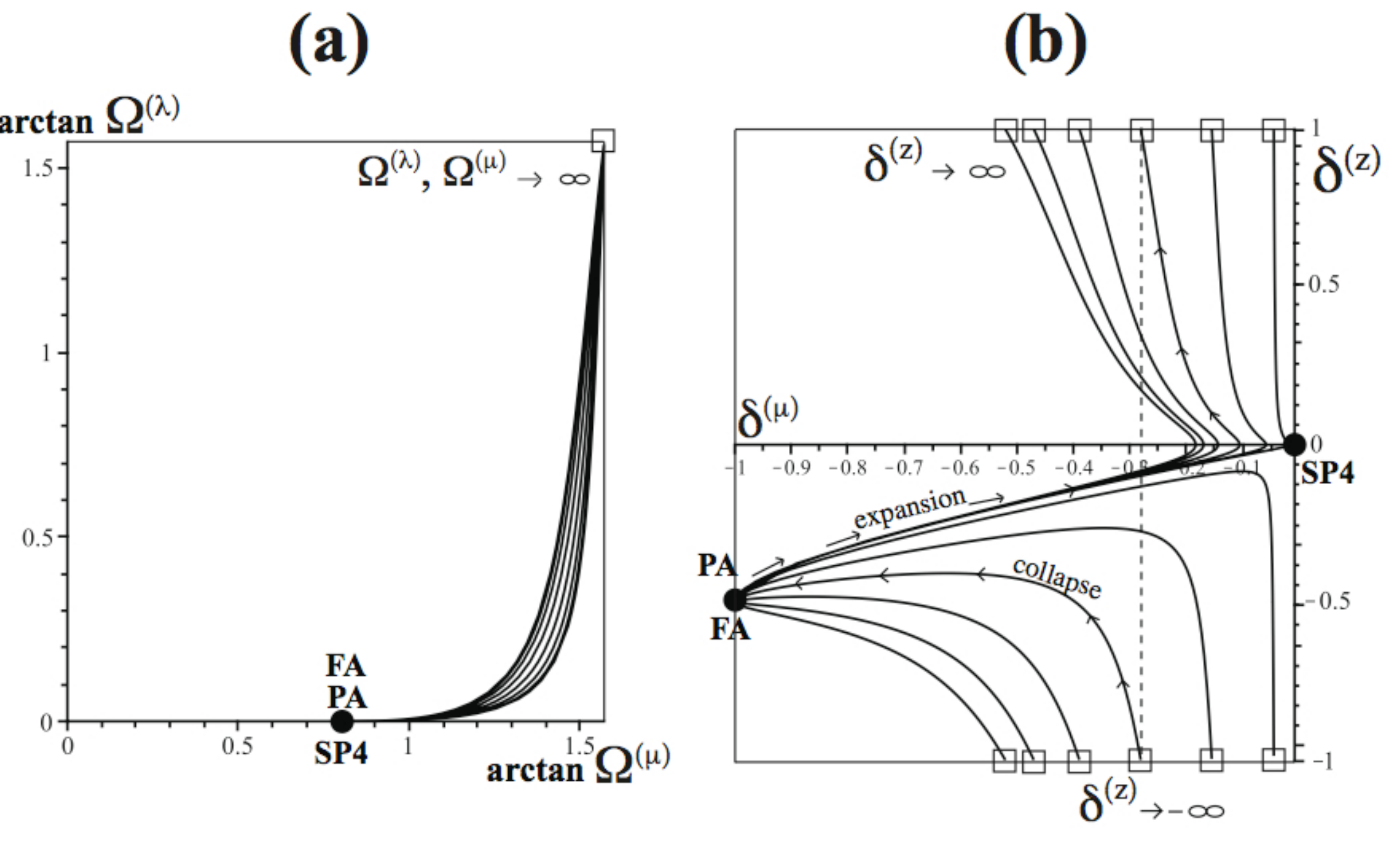}
\caption{{\bf All layers collapse.} This configuration has space slices with closed topology and follows from initial conditions (\ref{IVF_24a})--(\ref{IVF_24c}). See the text for further explanation on the phase space evolution.}
\label{fig7}
\end{center}
\end{figure}
%
\subsubsection{Collapsing models: kinematic pattern (ii).}

If we choose $\mu_{qi}\approx \kappa_{qi}>1$ with $\lambda\ll \mu_{qi},\kappa_{qi}$ for all $r$, we obtain strongly $\mu$--dominated $\Lambda$--LTB configurations that are similar to the case of pure dust $\Lambda=0$ with positive curvature in which layers expand from an initial curvature singularity (big bang), bounce and collapse to a second singularity (big crunch). Notice that, in general, neither singularity coincides with a simultaneous hypersurface. The space slices can have closed, open or wormhole topologies. An example of initial conditions leading to this type of  evolution is given by choosing $f=\sin r$ (closed topology) together with
\bse\ba \fl \mu_{qi}=m_{10}+\frac{m_{11}}{1+f^2(r)},\quad &
(m_{10}=0.3, \quad m_{11}=2.0), &\label{IVF_24a}\\
\fl \kappa_{qi}=k_{10}+ \frac{k_{11}}{1+
f^2(r)},&(k_{10}=0.01,\quad k_{11}=1.98),\quad &\label{IVF_24b}\\
\fl \lambda=0.01.&&\label{IVF_24c}\ea\ese
where $\kappa_{qi}$ complies with (\ref{grr1}) and (\ref{grr2}). Since dust layers can only evolve in the range $0<L<L_1$ (see section 6), the initial expansion reverses into a collapsing stage at a ``bouncing point'' $L=L_1(r)$ which is in general different for each $r$. At this bouncing point we have $z_q=0$, and so $\Omm,\,\Oml$ and $\Dz$ diverge, and as a consequence, we cannot find the collapsing stage of the trajectories (which lies beyond this point) with the numeric solutions of the dynamical system (\ref{DSa})--(\ref{DSd}). The full trajectories in this case must follow from the numeric solutions of the system (\ref{sys1a})--(\ref{sys1d}).

Phase trajectories are displayed in figure \ref{fig7}, where we have highlighted in panel (b) a representative trajectory marked by arrows (to facilitate the explanation). The homogeneous projection of the phase space evolution (panel (a)) is very simple: the expanding curves emerge from the past attractor {\bf \textrm{PA}}, reach the bouncing point where $z_q\to 0$ and $\Omm,\,\Oml$ diverge (marked by a square), the expansion reverses into a collapse and the curves return following the same trajectories (but run in reverse direction) towards {\bf \textrm{PA}}, which is now the a future attractor in the collapsing stage. In the inhomogeneous projection (panel (b), see representative curve), curves follow different paths in the expanding and collapsing stages. All curves initially expand and emerge from {\bf \textrm{PA}}, approach the saddle {\bf \textrm{SP4}}, and head upwards to the ``bouncing points'' marked by the upper squares where $z_q\to 0$ and $\Dz\to\infty$ for $\Dm$ variable. In the collapsing stage (see representative curve) the curves start in the bouncing points (lower squares) with $\Dz\to-\infty$ for $\Dm$ variable and evolve upwards to {\bf \textrm{PA}}, which is now a future attractor. The fact that expanding and collapsing trajectories are the same (with $\xi\to -\xi$) in the homogeneous projection (panel (a)), but are wholly different in the inhomogeneous projection (panel (b)) is an important inherent feature  due to the inhomogeneity of the models.

We can obtain the same type of collapsing configurations with slices having open and wormhole topologies by means of similar initial conditions, but with $f=\tan r,\,\sec r$.

\subsubsection{Bouncing models without collapse: kinematic pattern (iii).}

\begin{figure}[htbp]
\begin{center}
\includegraphics[width=4.0in]{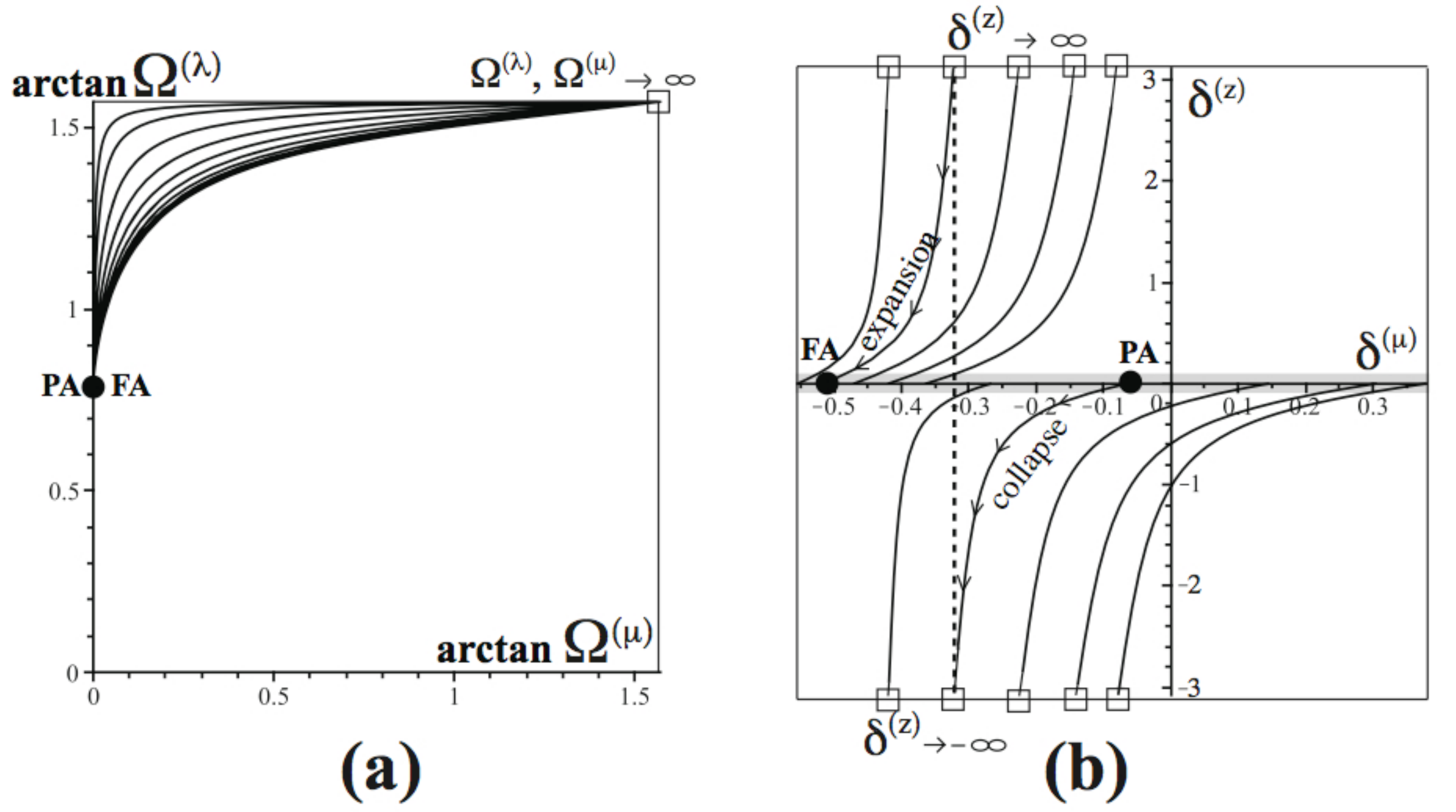}
\caption{{\bf All layers bounce.} This configuration follows from initial conditions (\ref{IVF_25a})--(\ref{IVF_25c}).}
\label{fig8}
\end{center}
\end{figure}

Strongly $\lambda$--dominated configurations follow by selecting $\mu_{qi}\ll \kappa_{qi}$ and $\mu_{qi}\ll\lambda$ for all $r$. Since all layers start collapsing from infinite $L$ but bounce and avoid collapsing to a singularity, these configurations have no equivalent in the pure dust case ($\lambda=\Oml=0$). The space slices can have closed, open or wormhole topologies. An example of initial conditions leading to this type of evolution follows from choosing $f=\tan r$ (open topology) and
\bse\ba \fl \mu_{qi}=m_{10}+\frac{m_{11}}{1+f^2(r)},\quad &
(m_{10}=0.01, \quad m_{11}=1.5), &\label{IVF_25a}\\
\fl \kappa_{qi}=k_{10}+\frac{k_{11}}{1+
f^2(r)},&(k_{10}=10.0,\quad k_{11}=30.0),\quad &\label{IVF_25b}\\
\fl \lambda=31.0.&&\label{IVF_25c}\ea\ese
The phase space trajectories are displayed by figure \ref{fig8}. Since dust layers can only evolve in the range $L>L_2$ (see section 6), the initial collapse from infinity bounces into an expansion at $L=L_2(r)$ which is in general different for each $r$. As in the collapsing case, at this bouncing point we have $z_q=0$, and so $\Omm,\,\Oml$ and $\Dz$ diverge, hence we have plotted the arctan of these variables in figure \ref{fig8}. As in the previous case, we have used the system (\ref{sys1a})--(\ref{sys1d}) to obtain the collapsing stage of the evolution.

The curve of past attractors {\bf \textrm{PA}} is given by (\ref{FA}), and is the same as the curve of future attractors for the perpetually expanding configurations of figures \ref{fig4}, \ref{fig5} and \ref{fig6}. Every phase space trajectory starts from a point (marked by {\bf \textrm{PA}}) in this curve of past attractors (thick gray line) and terminates in a different point in the same curve (of future attractors, marked as {\bf \textrm{FA}}). These attractors are projected into the same point in the homogeneous projection (panel (a)) and into two distinct points in the thick gray line in the homogeneous projection (panel (b)). In panel (a) the initially collapsing trajectories start at {\bf \textrm{PA}}, evolve upwards to the bounce points as $z_q\to 0$ and $\Omm,\Oml\to\infty$ (squares) and return to the same point (now a future attractor {\bf \textrm{FA}}) following the same trajectory. In panel (b) (see representative trajectory marked by arrows), trajectories start in a point {\bf \textrm{PA}} of the curve of attractors (thick gray line) and evolve downwards to the bounce points where $z_q\to 0$ and $\Dz\to-\infty$ (lower squares) and $\Dm$ variable. In the expanding phase trajectories start at a bounce points $z_q\to 0$ and $\Dz\to\infty$ (upper squares) and go downward towards the point {\bf \textrm{FA}} of the curve of attractors.  As in the previous case (collapsing layers), we also have a time symmetric reflection ($\xi\to-\xi$) in the collapsing/expanding evolution in the homogeneous variables, but not in the inhomogeneous variables.

Bouncing configurations that avoid collapse can be constructed with space slices having closed and wormhole topologies by means of similar initial conditions, but with $f=\sin r,\,\sec r$ and $\kappa_{qi}$ complying with (\ref{grr1}) and (\ref{grr2}). Their phase space trajectories are qualitatively similar to those of figure \ref{fig8}.

\subsubsection{Loitering models: kinematic pattern (iv).}

\begin{figure}[htbp]
\begin{center}
\includegraphics[width=4.0in]{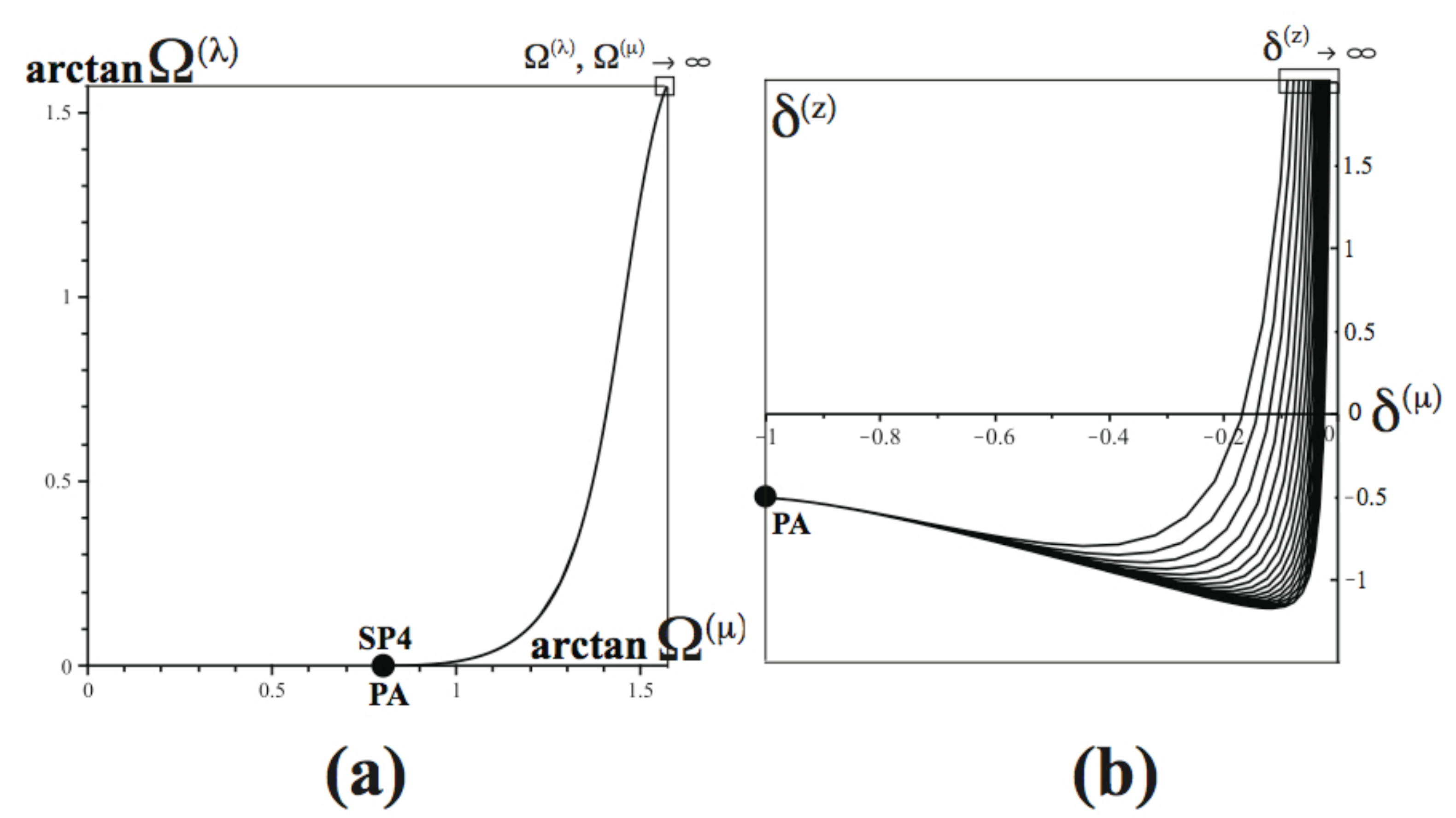}
\caption{{\bf Loitering layers.} This configuration corresponds to initial conditions complying with $\Qmin=0$.}
\label{fig9}
\end{center}
\end{figure}

Initial conditions for such models follow from taking $\kappa_{qi}>0$ and restricting the initial value functions so that $\Qmin = 0$ holds. Under these restrictions $\Lambda$--LTB models evolve towards an asymptotic static state, which is very unstable because of the extreme sensitivity to initial conditions exactly complying with $\Qmin=0$ (see section 6 and figure \ref{fig1}). As a consequence, the numeric integration of (\ref{DSa})--(\ref{DSd}) for these initial conditions is extremely susceptible to numerical errors. However, if $\Qmin = 0$ the  Friedman--like equation (\ref{sqHq}) has the following closed analytic solution:
\begin{equation} \fl L = \gamma_i\left\{2\tanh^2\left[\left(\frac{\gamma_i\,\lambda}{2}\right)^{1/2}\,(t-t_i)-\hbox{arctanh}\left(\frac{1}{\sqrt{2}}\left(1+\frac{1}{\gamma_i}\right)\right)\right]-1\right\},\label{Lloit}\end{equation}
where $\gamma_i^2=\kappa_{qi}/(3\lambda)$ and we eliminated $\mu_{qi}$ from (\ref{Qmindef}). Since all phase space variables can the computed analytically from (\ref{Lloit}) by means of (\ref{LGdef})--(\ref{Oml}), we can plot the phase space trajectories given a choice of initial conditions.  Considering $\gamma_i =(2+r^4)/(1+r^4)$ and $\lambda=0.1$, we obtain the phase space diagrams displayed by figure \ref{fig9}. All trajectories in both panels (a) and (b) begin at the past attractor {\bf \textrm{PA}} (big bang) and evolve towards the static state characterized by $z_q\to 0$, which implies  $\Omm,\Oml,\Dz\to\infty$. 
%

\subsection{Configurations with mixed kinematic patterns: kinematic patterns (v).}

Since initial value functions depend on $r$, we can select initial conditions so that different kinematic patters occur in the same configuration. In practical terms, such initial conditions must comply with (\ref{Qmin}) and would yield curves $[\Ommi(r),\Omli(r)]$ that traverse the shaded and non--shaded areas of figure \ref{fig1}b. Notice that these ``mixed'' pattern configurations can be easily constructed within a single inhomogeneous model, but cannot be set up with any single homogeneous model.

\begin{figure}[htbp]
\begin{center}
\includegraphics[width=4in]{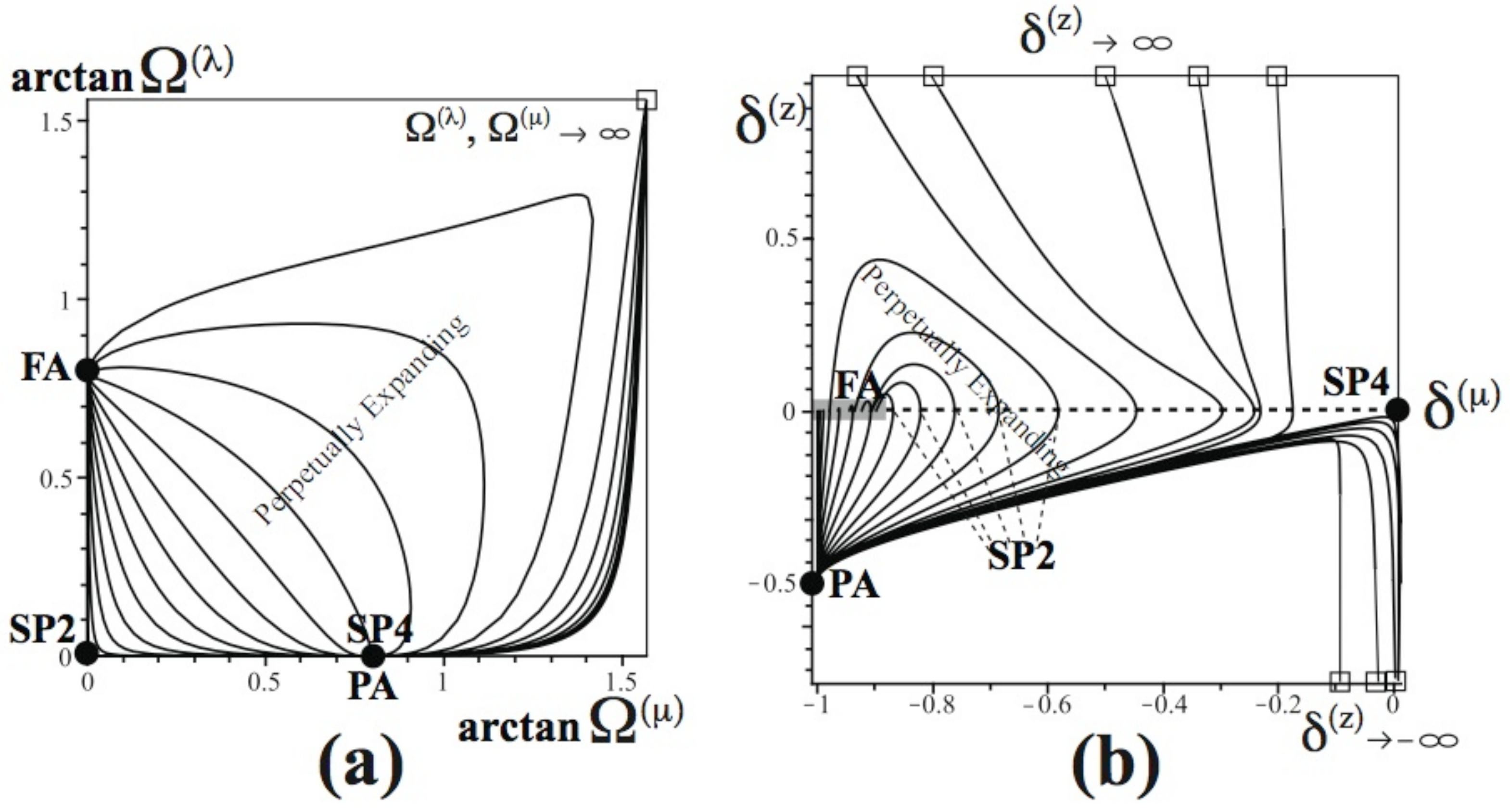}
\caption{{\bf Structure formation scenario.} Initial conditions for this configuration are given by (\ref{IVF_27a})--(\ref{IVF_27c}). }
\label{fig10}
\end{center}
\end{figure}
\begin{figure}[htbp]
\begin{center}
\includegraphics[width=4.0in]{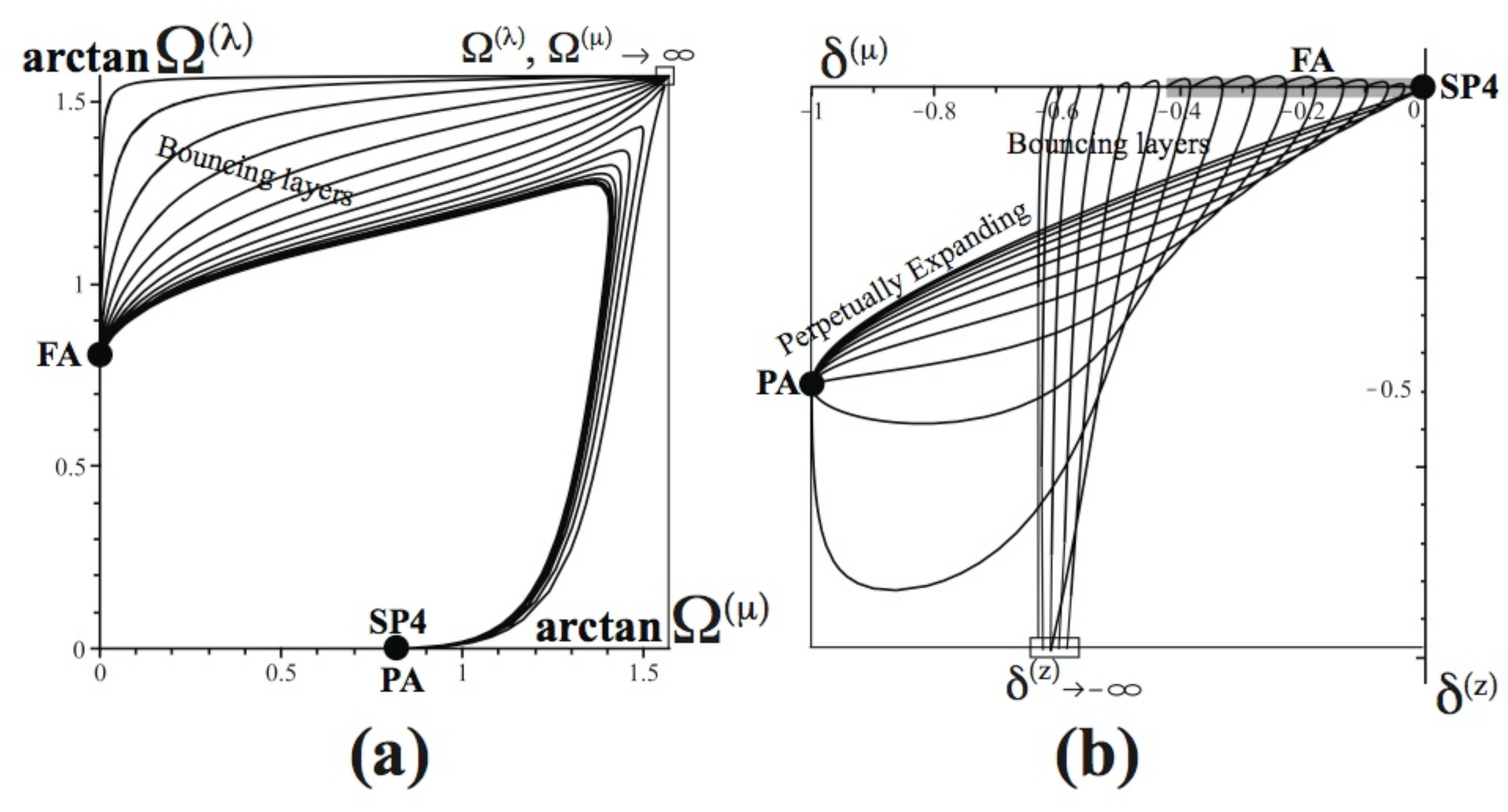}
\caption{{\bf Mixed expansion and bounce.} Initial conditions are given by (\ref{IVF_28a})--(\ref{IVF_28c}).}
\label{fig11}
\end{center}
\end{figure}
%
\subsubsection{Structure formation scenario.}

Interesting configurations that can be associated with ``structure formation'' toy models  follows from initial conditions in which (\ref{Qmin}) holds (shaded $\mu$--dominated region in figure \ref{fig1}b) in a radial range containing a symmetry center ($0\leq r\leq r_b$) and fails to hold (non--shaded region in figure \ref{fig1}b) for all $r>r_b$. This  corresponds to an inner spherical region undergoing local collapse surrounded by an expanding ``background''.  While this structure formation scenario can also occur with pure dust solutions, in this case the boundary $r=r_b$ between the collapsing region and the expanding background corresponds to the worldline where spatial curvature passes from positive to negative or zero. However, if $\lambda>0$ this scenario can also occur with all layers having positive spatial curvature, as a perpetually expanding background with positive curvature is possible.

An example of initial conditions leading to the structure formation toy model follows by choosing $f(r)=\tan r$ (open topology) with
\bse\ba \fl \mu_{qi}=m_{10}+\frac{m_{11}-m_{10}}{1+f^3(r)},\quad &
(m_{10}=0.0, \quad m_{11}=20.1), &\label{IVF_27a}\\
\fl \kappa_{qi}=k_{10}+\frac{k_{11}-k_{10}}{1+
f^4(r)},&(k_{10}=-4.1,\quad k_{11}=35.5),\quad &\label{IVF_27b}\\
\fl \lambda=0.75.&&\label{IVF_27c}\ea\ese
The phase space trajectories are displayed by figure \ref{fig10}. All curves start their evolution at the past attractor {\bf \textrm{PA}} marking the initial singularity (big bang). However, the ``inner'' layers around the center ($r=0$) and the ``outer'' layers $r>r_b$ follow a different phase space evolution.

Since outer dust layers for a given $r>r_b$ expand perpetually, their evolution is analogous to that of perpetually expanding configurations depicted by figures \ref{fig4}, \ref{fig5} or \ref{fig6} (depending on the spatial curvature and the asymptotics of $\mu_{qi}$ and $\kappa_{qi}$). In the particular example we are considering, these trajectories (see figure \ref{fig10}) have analogous forms to those of these three figures. From the past attractor {\bf \textrm{PA}} they approach the saddle {\bf \textrm{SP2}} and end in the line of future attractors {\bf \textrm{FA}} (whose inhomogeneous projection is marked as a thick gray line in panel (b)).

The inner layers, on the other hand,  evolve in a similar way as the trajectories of collapsing configurations displayed  by figure \ref{fig7}, they initially expand away from the past attractor {\bf \textrm{PA}} approach the saddle {\bf \textrm{SP4}} and go upwards to the bounce points as $z_q\to 0$, with $\Omm,\,\Oml$ and $\Dz$ diverging (the bounce points are marked by small squares in both panels).  The main difference with the ``pure'' collapse case of figure \ref{fig7} is that the trajectories of inner layers closer to the outer perpetually expanding layers reach their bouncing points (see panel (b)) as $\Dz\to\infty$, whereas inner layers closer to the center do so as $\Dz\to-\infty$. The collapsing stage of these layers is not shown in figure \ref{fig10}, but it can be readily obtained from system (\ref{sys1a})--(\ref{sys1d}): it is analogous to that depicted in figure \ref{fig7}: the collapsing trajectories in panel (a) are simply the time reversal of the expanding ones, whereas in panel (b) they start from a bouncing point towards a future attractor that exactly coincides with {\bf \textrm{PA}} (the collapsing singularity) but following a different path to that of the expanding curves.

\subsubsection{Bouncing/expanding models.}

Another mixed kinematics configuration follows if dust layers around the center expand from an initial singularity, while external layers collapse from infinity and bounce back. In this case initial conditions need to comply with $\Ommi\approx\Omli$ near the center passing to $\Omli\gg\Ommi$ as $r$ grows, so that the curve $[\Ommi(r),\Omli(r)]$ would be passing from the non--shaded to area of figure \ref{fig1}b towards the $\lambda$--dominated shaded region. As an example of such evolution pattern we use initial conditions
\bse\ba \fl \mu_{qi}=m_{10}+\frac{m_{11}-m_{10}}{1+f^2(r)},\quad &
(m_{10}=0.01, \quad m_{11}=15.0), &\label{IVF_28a}\\
\fl \kappa_{qi}=k_{10}+\frac{k_{11}-k_{10}}{1+
f^2(r)},&(k_{10}=10.0,\quad k_{11}=30.0),\quad &\label{IVF_28b}\\
\fl \lambda=10.0,&&\label{IVF_28c}\ea\ese
with $f=\tan r$, leading to the phase space diagrams of figure \ref{fig11} that clearly show the mixed evolution patterns. Layers expanding perpetually from an initial singularity behave as the curves in figure \ref{fig6}: they start from the past attractor {\bf \textrm{PA}}, approach the saddle {\bf \textrm{SP4}} and terminate in the future attractor {\bf \textrm{PA}}. The bouncing curves are analogous to those of figure \ref{fig8}: they originate from the future attractor {\bf \textrm{PA}} (a past attractor for these curves) and evolve towards the bounce $z_q\to 0$ (or $\Omm,\,\Oml \to\infty,\,\Dz\to -\infty$ with $\Dm$ variable), and then expand back to {\bf \textrm{PA}} (now a future attractor). We only display in figure \ref{fig11} the collapsing stage of this evolution, as the curves in the expanding stage are qualitatively analogous to the expanding curves of figure \ref{fig8}.


\section{Conclusion.}

We have conducted a comprehensive and detailed dynamical systems study of $\Lambda$--LTB spacetimes by means of covariant quasi--local (QL) variables, generalizing a previous dynamical systems study \cite{suss08} on LTB dust models with $\Lambda=0$. Since the phase space for the dynamical system (\ref{DSa})--(\ref{DSd}) is 4--dimensional, we have examined and plotted phase space trajectories in two 2--dimensional subsets: the ``homogeneous'' and ``inhomogeneous'' subspaces, $\textrm{H}_0$ and $\textrm{I}$, defined by the projections (\ref{Hproj}) and (\ref{Iproj}), and respectively displayed in panels (a) and (b) of figures \ref{fig4}--\ref{fig11}. The decomposition of the phase space into $\textrm{H}_0$ and $\textrm{I}$ provides an important theoretical connection between the dynamical system study of $\Lambda$--LTB spacetimes and the perturbation formalism presented in \cite{sussPRD,sussLTB1}. In this formalism, the fluctuations $\{\Dm,\,\Dz\}$ (which parametrize $\textrm{I}$) are rigorously characterized as covariant and gauge invariant non--linear perturbations on a FLRW background defined by quasi--local scalars, which satisfy FLRW dynamics, and hence corresponding to the projected homogeneous subspace $\textrm{H}_0$ parametrized by $\{\Omm,\,\Oml\}$.

In LTB dust models with $\Lambda=0$ all configurations in which the space slices $\T[t]$ have closed ($\mathbb{S}^3$) or wormhole ($\mathbb{S}^3\times \mathbb{R}$ or $\mathbb{S}^3\times \mathbb{S}^1$) topologies, the regularity conditions (\ref{grr1})--(\ref{grr2}) imply that spatial curvature must be positive and thus, perpetual expansion is not possible \cite{sussLTB1,ltbstuff,suss02}. However, this restriction is no longer true when $\Lambda>0$. While (\ref{grr1})--(\ref{grr2}) still requires positive spatial curvature for configurations with these topologies, positive spatial curvature no longer forbids perpetual expansion (see section 7). Hence, perpetually expanding and fully regular $\Lambda$--LTB models with closed or wormhole topologies are possible.

The homogeneous projections in panels (a) of figures \ref{fig4}--\ref{fig11} show that phase space trajectories are qualitative analogous to those of dust--$\Lambda$ FLRW models. However, each FLRW spacetime configuration would correspond to one (and only one) phase space trajectory in a $\{\Omm,\,\Oml\}$ diagram (because each spacetime configuration is uniquely determined by initial conditions given by constant $\Ommi,\Omli$). As a contrast, since initial conditions for any single $\Lambda$--LTB configuration depend on $r$, all displayed curves in the diagrams in the panels (a) of each one of these figures  corresponds to a single spacetime configuration. In other words, in terms of the projection (\ref{Hproj}) each $\Lambda$--LTB configuration is formally equivalent to a superposed one--parameter family of dust--$\Lambda$ FLRW models. 

In all $\Lambda$--LTB configurations (except those like the one in figure \ref{fig8}) the phase space trajectories initiate in the same source or past attractor {\bf{PA}}, defined in (\ref{PA}) and associated with an initial singularity. This critical point can be projected into the past attractor of the homogeneous subspace, and can be identified with a spatially flat FLRW dust cosmology ($\Omm=1,\,\Oml=0$). Since $\Dm=\Dz=-1$, this critical point exactly coincides with the past attractors of LTB dust solutions with $\Lambda=0$ and spatially flat models with $\Lambda>0$, thus indicating that the effects of $\Lambda$ and of spatial curvature are negligible in the early stages of the evolution near the big bang. The past attractor {\bf{PA}} can also be identified with self--similar conditions \cite{suss08}, which is consistent with the fact that both spatially flat FLRW and LTB dust spacetimes with $\Lambda=0$ are compatible with a homothetic Killing vector. The past attractor (\ref{PA}) becomes the future attractor or sink for phase space trajectories of layers that collapse to a curvature singularity (big crunch). This is so, either for a collapse that is a trivial time reversal of a perpetual expansion (figures \ref{fig4}--\ref{fig6}), or for configurations where all layers expand and collapse (figure \ref{fig7}), or with mixed dynamics in which some layers expand perpetually and some collapse (figure \ref{fig10}).

Phase space trajectories of perpetually expanding layers ($z_q>0$) terminate in the same future attractor (sink) {\bf{FA}}, defined in (\ref{FA}): figures \ref{fig4}--\ref{fig6} and in the perpetually expanding trajectories of figures \ref{fig8}, \ref{fig10} and \ref{fig11}. This critical point can be projected into the future attractor of the homogeneous subspace ($\Omm=0,\,\Oml=1$), and is also the future attractor for spatially flat configurations. Hence, it can be identified with de Sitter spacetime and spatially flat conditions, indicating that the effects of $\Lambda$ are dominant in the future late stages of the evolution of $\Lambda$--LTB configurations in which the worldlines of expanding dust layers are infinitely inextensible. As a consequence, with the exception of the unstable loitering models, all $\Lambda$--LTB models endowed with a late time evolution are compatible with the ``{\it cosmic no hair}'' conjecture~\cite{barrow,cnh}, with the asymptotic de Sitter state identified with the critical point {\bf{FA}}. For dust layers collapsing from infinity ($L\to\infty$), the phase space trajectories emerge from a past attractor which exactly coincides with the point {\bf{FA}}.

Seen from the inhomogeneous projection (\ref{Iproj}) (panels (b)), the evolution towards {\bf{FA}} of phase space trajectories of perpetually expanding layers contains extra information not available in the projection (\ref{Hproj}). Trajectories of layers with negative spatial curvature (figures \ref{fig4}--\ref{fig5}) approach the Minkowskian saddle {\bf{SP2}}, whereas layers with positive curvature approach the saddle {\bf{SP4}} (figures \ref{fig6}, \ref{fig7}, \ref{fig10} and \ref{fig11}) associated with spatially flat conditions. The approach to {\bf{SP2}} clearly indicates an intermediate low density state, and also occurs in trajectories of the FLRW phase space (notice that {\bf{SP2}} is projected by (\ref{Hproj}) into the single saddle of this phase space). However, the approach to {\bf{SP4}} has no equivalent in trajectories of the FLRW phase space of layers evolving towards the future attractor.

In configurations containing dust layers that bounce and collapse ($z_q$ changes sign: figures \ref{fig7}--\ref{fig11}) we used the system (\ref{sys1a})--(\ref{sys1d}) to obtain the full phase space trajectories. In all cases the phase space trajectories in the homogeneous projection  start in the past attractor {\bf{FA}}, reach infinite values $\Omm,\Oml\to\infty$ as $z_q=0$ and return along the same trajectories to the same attractor (which is now a future attractor). However, in the inhomogeneous projection (panels (b) of figures \ref{fig7} and \ref{fig8}) the curves do not return to the attractor along the same trajectories: if the pattern is expansion/collapse (figure \ref{fig7}) they reach $\Dz\to\infty$ as $z_q=0$ and return to the attractor from $\Dz\to-\infty$. If the pattern is collapse/bounce (figure \ref{fig8}), the trajectories reach $\Dz\to-\infty$ and return to the attractor from $\Dz\to\infty$. The same effect would happen in the mixed pattern configurations displayed in figures \ref{fig10} and \ref{fig11}, but was not included in the plots as the full curves would make a very messy pattern. This difference in the behavior of curves emerging/returning from/to an attractor is an effect inherent in the inhomogeneity of LTB solutions.

The bouncing and loitering models deserve a separate mention, as they have no equivalence in the case $\Lambda=0$. In the bouncing models (figure \ref{fig8}), layers initially collapse from infinity, hence their past attractor coincides with the future attractor of perpetually expanding configurations (figures  \ref{fig4}--\ref{fig6}).  In the loitering configuration that we examined (figure \ref{fig9}), the phase space trajectories emerge from the past attractor {\bf{PA}}, but as the layers become asymptotically static $z_q\to 0$ the trajectories evolve towards $\Omm,\Oml,\Dz\to\infty$. In the homogeneous projection (panel (a) of figure \ref{fig9}) the trajectories are indistinguishable from those of figure \ref{fig7}, but in the inhomogeneous projection (panel (b) of figure \ref{fig9}) the curves reach $\Dz\to\infty$ along different paths and converge towards $\Dm\to 0$. This effect is an inherent feature of inhomogeneity and does not occur in FLRW phase space diagrams for loitering models. It is worthwhile remarking that loitering models do not evolve towards an asymptotic de Sitter state associated with the future attractor {\bf{FA}}.   

The dynamics of $\Lambda$--LTB models is not much different (qualitatively speaking) from that of the homogeneous FLRW models that they generalize. Essentially, the past attractor is a state close to spatially flat FLRW and self--similarity, while the future attractor is a de Sitter state dominated by $\Lambda$ (in agreement with the ``cosmic no hair'' conjecture). The two most significant effects that arise from the inhomogeneity of $\Lambda$--LTB models are:
\begin{enumerate}
\item  the possibility of accommodating wholly different kinematic patterns in the same spacetime configuration (see figures \ref{fig10}--\ref{fig11}). However, even when the kinematic patterns of all layers are qualitatively similar, each unique $\Lambda$--LTB spacetime can be understood as an inhomogeneous model made up of some sort of superposed FLRW models (as the phase space trajectory of each comoving worldline can be projected by (\ref{Hproj}) into the phase space trajectory of a unique FLRW model).
\item the asymmetry in the evolutions from a past attractor and a bouncing point and from that point to the same attractor, which becomes a future attractor (see panels (b) of figures \ref{fig7} and \ref{fig8})
\end{enumerate}
However, in spite of the simplification involved in assuming spherical symmetry and geodesic motion, the present dynamical system study clearly shows that even this idealized level of inhomogeneity does provide extra degrees of freedom that can be be very handy for constructing useful toy models of astrophysical and cosmological inhomogeneities.

\section*{Acknowledgments.}

The authors acknowledge financial support from grant DGAPA--PAPIIT--UNAM 119309. G.I. research was funded by the Conseil R\'egional de Bourgogne.

\end{document}